\documentclass[aps,prl,reprint,amsmath,amssymb,superscriptaddress]{revtex4-2}
\usepackage{amsmath} %\text{} for text in math
\usepackage{bm}  %\bm{} for bold font in math
\usepackage{graphicx}
\usepackage{physics}
\usepackage{color}
\usepackage[colorlinks=true, citecolor=blue, linkcolor=blue, urlcolor=blue]{hyperref}
\graphicspath{ {./}{./figures/} }

\newcommand{\rr}{\mathbf{r}}
\newcommand{\RR}{\mathbf{R}}
\newcommand{\kk}{\bm{k}}
\renewcommand{\qq}{\bm{q}}

\begin{document}

\title{\textit{Ab initio} electron-phonon interactions in correlated electron systems} %from Hubbard-corrected density functional theory}
%{Electron-phonon interactions in a Mott insulator in the framework of DFT+$U$}
\author{Jin-Jian Zhou}
\thanks{J.-J.Z. and J.P. contributed equally to this work}
\author{Jinsoo Park}
\thanks{J.-J.Z. and J.P. contributed equally to this work}
\affiliation{Department of Applied Physics and Materials Science, California Institute of Technology, Pasadena, CA 91125, USA.}

% Iurii Timrov
\author{Iurii Timrov}
\affiliation{Theory and Simulation of Materials (THEOS) and National Centre for Computational Design and Discovery of Novel Materials (MARVEL), École Polytechnique Fédérale de Lausanne (EPFL), CH-1015 Lausanne, Switzerland}

% Andrea Floris
\author{Andrea Floris}
\affiliation{School of Chemistry, University of Lincoln, Brayford Pool, Lincoln LN6 7TS, United Kingdom}

% Matteo Cococcioni
\author{Matteo Cococcioni}
\affiliation{Department of Physics, University of Pavia, Via A. Bassi 6, I-27100 Pavia, Italy}

% Nicola Marzari
\author{Nicola Marzari}
\affiliation{Theory and Simulation of Materials (THEOS) and National Centre for Computational Design and Discovery of Novel Materials (MARVEL), École Polytechnique Fédérale de Lausanne (EPFL), CH-1015 Lausanne, Switzerland}

\author{Marco Bernardi}
\email[Corresponding author: ]{bmarco@caltech.edu}
\affiliation{Department of Applied Physics and Materials Science, California Institute of Technology, Pasadena, CA 91125, USA.}

%\date{\today}

\begin{abstract}
Electron-phonon ($e$-ph) interactions are pervasive in condensed matter, governing phenomena such as transport, superconductivity, charge-density waves, polarons and metal-insulator transitions.
% . However, studies of $e$-ph interactions in
First-principles approaches enable accurate calculations of $e$-ph interactions in a wide range of solids. However, they remain an open challenge in correlated electron systems (CES), where density functional theory often fails to describe the ground state. Therefore reliable $e$-ph calculations remain out of reach for many transition metal oxides, high-temperature superconductors, Mott insulators, planetary materials and multiferroics. 
% typically Unfortunately most of first-principles approaches result quite inaccurate on evaluating the e-ph
%interactions on important systems due to their failure in capturing electronic correlations.
%Here we present the first quantitatively predictive calculations of el-ph couplings for a correlated
%electron system (CES). Calculations are performed 
%Reliable $e$-ph calculations are crucial in transition metal oxides, high-temperature superconductors, Mott insulators and multiferroics. 
%
Here we show first-principles calculations of $e$-ph interactions in CES, using the framework of Hubbard-corrected density functional theory (DFT+\textit{U}) and its linear response extension (DFPT+\textit{U}), which can describe the electronic structure and lattice dynamics of many CES. 
%an ab-initio framework enabling accurate e-ph calculations in correlated materials, based on Hubbard-corrected DFT+$U$ and density functional perturbation theory (DFPT+U). 
We showcase the accuracy of this approach for a prototypical Mott system, CoO, carrying out a detailed investigation of its $e$-ph interactions and electron spectral functions. While standard DFPT gives unphysically divergent and short-ranged $e$-ph interactions, DFPT+\textit{U} is shown to remove the divergences and properly account for the long-range Fr\"ohlich interaction, allowing us to model polaron effects in a Mott insulator. 
%
%Our findings highlight the interplay of electron, spin and lattice degrees of freedom in CES and their combined effect on the $e$-ph coupling. 
Our work establishes a broadly applicable and affordable approach for quantitative studies of $e$-ph interactions in CES, a novel theoretical tool to interpret experiments in this broad class of materials.
\end{abstract}
\maketitle
%
%\section{Introduction}
%
%Transition metal oxides (TMOs) 
Strongly correlated materials are at the center of exciting advances in condensed matter physics.  
These correlated electron systems (CES) can host states of matter ranging from high-temperature superconductivity~\cite{pickettElectronic1989} to Mott transitions~\cite{mottBasis1949,mottMetalInsulator1968}, colossal magnetoresistance~\cite{ramirezColossal1997} and multiferroicity~\cite{schmidMultiferroic1994}. 
Electron-phonon ($e$-ph) interactions play an important role in these phenomena, often governing their origin and temperature dependence. 
A promising direction to study quantitatively $e$-ph interactions in CES is using first-principles calculations, %~\cite{giustinoElectronphonon2007,bernardiFirstprinciples2016,agapitoInitio2018,zhouInitio2016}.
%In the typical approach, 
where one employs density functional theory (DFT) to compute the electronic structure, density functional perturbation theory~\cite{baroniPhonons2001} (DFPT) for the lattice dynamics, and their combination to obtain the $e$-ph coupling~\cite{bernardiFirstprinciples2016,agapitoInitio2018}. 
% giustinoElectronphonon2007,
%\\
%\indent
This approach can successfully describe $e$-ph interactions and electron dynamics in a wide range of materials~\cite{bernardiInitio2014,bernardiFirstprinciples2016,  zhouInitio2016, agapitoInitio2018, zhouElectronPhonon2018,zhouPredicting2019,Park-PRL2007, Floris2007,Sjakste, liElectrical2015,liuFirstprinciples2017,maFirstprinciples2018,Ponce-PRB, parkSpinphonon2020}. 
\\
\indent
% charge transport and carrier dynamics
However, computing $e$-ph interactions in CES remains challenging as standard DFT usually fails to describe their ground state, mainly due to self-interaction errors in open subshells of localized~\textit{d} or $f$ electrons. 
% localized
In addition, correlated transition metal oxides (TMOs) often exhibit strong $e$-ph coupling and polaron effects, requiring treatments beyond lowest-order perturbation theory~\cite{zhouPredicting2019}. 
% 
%, mainly due to its charge-density based representation of exchange and correlation interactions.
%
% Intro DFT+U  
Widely used first-principles approaches to compute the ground state of CES include DFT+\textit{U}~\cite{anisimovBand1991,anisimovChargeordered1996,anisimovFirstprinciples1997,dudarevElectronenergyloss1998}, hybrid functionals~\cite{heydHybrid2003}, and dynamical mean-field theory~\cite{georgesDynamical1996,kotliarElectronic2006}. Developing accurate $e$-ph calculation in any of these frameworks is an important open challenge $-$ if fulfilled, it would advance investigations of the rich physics of CES and significantly expand the scope of first-principles studies of $e$-ph interactions.
\\
\indent
The DFT+\textit{U} method ~\cite{anisimovBand1991,anisimovChargeordered1996,anisimovFirstprinciples1997,dudarevElectronenergyloss1998} is particularly promising to mitigate the self-interaction error of DFT, 
% improve the treatment
using the Hubbard correction to better capture the physics of localized \textit{d} electrons~\cite{Kulik2006,Kulik2015}. %
%
%such as phonon dispersions~\cite{florisVibrational2011,florisHubbardcorrected2020a} and Raman spectra~\cite{Blanchard2014,miwaPrediction2018}. 
%
It can predict the ground state of various families of correlated TMOs, including Mott insulators~\cite{anisimovBand1991}, high-temperature superconductors~\cite{pesantDFT2011} and multiferroics~\cite{baettigFirst2005,rondinelliNon2009}. 
% and rare earth compounds with localized $f$ and Raman spectra 
Its linear response variant, DFPT+\textit{U}, has been employed successfully to study the lattice dynamics of TMOs~\cite{florisVibrational2011,florisHubbardcorrected2020a,Blanchard2014,miwaPrediction2018}. As the Hubbard-\textit{U} value can be computed \textit{ab initio}~\cite{timrovHubbard2018}, as we do here, the framework is entirely free of empirical parameters.
%Leveraging recent advances in DFPT+\textit{U}~\cite{florisHubbardcorrected2020a}, reliable software for $e$-ph calculations~\cite{zhouPerturbo2020}, and numerical treatments of strong $e$-ph interactions in TMOs~\cite{zhouPredicting2019}. 
\\  
\indent
%This scenario points to a key development needed 
%to expand the scope of first-principles studies of $e$-ph interactions $-$ the inclusion of the Hubbard \textit{U} correction, 
%
% HERE WE SHOW 
%
In this Letter, we show calculations of $e$-ph interactions in the framework of DFT+\textit{U}, focusing on a prototypical Mott insulator, cobalt oxide (CoO), as a case study. 
%Our method is free of adjustable parameters, including the Hubbard~\textit{U} term, which is determined \textit{ab initio} with a linear response approach~\cite{timrovHubbard2018}. 
%%Application of this approach to a prototypical TMO, cobalt oxide (CoO), demonstrates the dramatic impact of the Hubbard~\textit{U} correction on the $e$-ph interactions in TMOs. 
% specific electronic structure, phonon dispersion, and DFT+U
%Although the ground state and $e$-ph interactions in CoO are entirely incorrect in DFT, 
%The dominant Fr\"ohlich $e$-ph interaction is correctly captured in DFT+\textit{U} correcting the localization of the 3\textit{d} bands and
While DFT predicts CoO to be a dynamically unstable metal,  DFT+\textit{U} correctly predicts its antiferromagnetic insulating ground state~\cite{florisHubbardcorrected2020a}.  
We thus find that the long-range Fr\"ohlich $e$-ph interaction is restored in DFPT+\textit{U}, and unphysical divergences of the $e$-ph coupling due to spurious soft modes are removed. %~\cite{florisHubbardcorrected2020a,zhouElectronPhonon2018}. 
%and the resulting divergences of the $e$-ph coupling. 
%by stabilizing the correct antiferromagnetic ground state.  
%, which become well-behaved throughout the Brillouin Zone (BZ).
With the correct Fr\"ohlich interaction in hand, we study the electron spectral function with a cumulant approach, revealing the formation of a polaron state with sharp quasiparticle and satellite peaks at low temperature that broaden and disappear entirely at room temperature. 
The Hubbard~\textit{U}-derived $e$-ph perturbation, missing in DFPT, is found to act primarily on the partially filled \textit{d} bands of each spin channel, showing the impact of the $d$ electron Coulomb repulsion on $e$-ph interactions.  
The DFT+\textit{U} $e$-ph calculations developed in this work are poised to advance the understanding of $e$-ph coupling, transport and superconductivity in strongly correlated materials.
\\ %various families of 
\indent
%\section{Methods}
% part 1:  electron-phonon coupling and e-ph matrix elements, 
% g = <||>, similar to the first part in my 2018 PRL paper. 
For quantitative studies of $e$-ph interactions, of key interest are the $e$-ph matrix elements, $g_{mn\nu}^\sigma(\kk,\qq)$, which quantify the probability amplitude for an electron in a Bloch state $\ket{\psi_{n\kk\sigma}}$, with band index $n$, spin $\sigma$ and crystal momentum $\kk$, to scatter into a final state $\ket{\psi_{m\kk+\qq\sigma}}$ %with the same spin 
by emitting or absorbing a phonon with mode index $\nu$, wave-vector $\qq$, energy $\hbar \omega_{\nu\qq}$, and displacement eigenvector $\bf{e}_{\nu\qq}$~\cite{bernardiFirstprinciples2016, zhouPerturbo2020,Giannozzi_2017},
% MATRIX ELEMENT (GENERAL) 
\begin{equation} \label{eq:me}
g_{mn\nu}^\sigma(\kk,\qq) \!=\! \left(\frac{\hbar}{2\omega_{\nu \qq}}\right)^{\!\frac{1}{2}} \!\sum_{I}\frac{\bf{e}_{\nu\qq}^{ \textit{I}}}{\sqrt{M_I}} \bra{\psi_{m\kk+\qq\sigma}} 
d_{\qq I} \hat{V}^\sigma\! \ket{\psi_{n\kk\sigma}}\!, 
\end{equation}
where $d_{\qq I} \hat{V}^\sigma \equiv \sum_{p} e^{i\qq \cdot \RR_p} d_{pI} \hat{V}^\sigma $ is the $e$-ph perturbation due to the change of the potential acting on an electron with spin $\sigma$ from a unit displacement of atom $I$ (with mass $M_I$ and located in the unit cell at $\RR_p$).
\\%with site index single-particle
\indent
%partr 2: in DFT+U framework, correction to the calculation of g
% 1, phonon frequencey and phonon eigen-displacement from DFPT+U
% 2, correction to the conventional dV_ks part,  we now have an additional dV_hub, introduce different part of the dV_hub. 
In DFPT+\textit{U}, besides the usual Kohn-Sham (KS) perturbation potential~\cite{zhouPerturbo2020}, %due to the phonon perturbation 
there is an additional term from the perturbation of the Hubbard potential~\cite{florisHubbardcorrected2020a}: %part
\begin{equation} \label{eq:dv}
d \hat{V}^\sigma=d \hat{V}^\sigma_{\text{KS}}+d \hat{V}^\sigma_{\text{Hub}}.
\end{equation}
This Hubbard perturbation potential is the sum of projector and occupation-matrix derivative terms~\cite{florisHubbardcorrected2020a}, 
\begin{equation} \label{eq:dvhub}
\begin{split}
d \hat{V}^\sigma_{\text{Hub}}=&
\sum_{I m_1 m_2}U^I \left( \frac{\delta_{m_1 m_2}}{2}-n_{m_1 m_2}^{I\sigma} \right)\partial\hat{P}_{m_1 m_2}^I \\
&-\sum_{I m_1 m_2}U^I (d n_{m_1 m_2}^{I\sigma} ) \hat{P}_{m_1 m_2}^I,
%V_U^\sigma \partial \hat{P} +  \hat{P}dV_U^\sigma.
%\sum_{I m_1 m_2}U^I (d n_{m_1 m_2}^{I\sigma} ) \hat{P}_{m_1 m_2}^I,
%\frac{\partial V_U^\sigma}{\partial\mu}\hat{P}
%\frac{\partial V_U}{\partial\mu} \hat{P}
\end{split}
\end{equation} %the corresponding 
where $m_1$ and $m_2$ are magnetic quantum numbers of the 3\textit{d} orbitals, $U^I$ is the effective Hubbard parameter for atom $I$, and $n_{m_1 m_2}^{I\sigma}$ is the occupation matrix for orbitals with magnetic quantum numbers $m_1$ and $m_2$ on atom $I$,
\begin{equation}
n_{m_1 m_2}^{I\sigma}=\sum_{n \kk}    \bra{\psi_{n\kk\sigma}}\hat{P}_{m_2 m_1}^I \ket{\psi_{n\kk\sigma}}.
\end{equation}
Here, $\hat{P}$ is the generalized projector on the space 
of the localized atomic orbitals $\varphi_m^I$,
\begin{equation}
\hat{P}_{m_2 m_1}^I=\hat{S}\ket{\varphi_{m_2}^I}\bra{\varphi_{m_1}^I}\hat{S},
\end{equation}
and $\hat{S}$ is the overlap operator of the ultrasoft pseudopotential~\cite{vanderbiltSoft1990}. 
%\\
%\indent 
%or the response of 
%independent of the change in the wave functions and occupation matrices, and is 
In Eq.~\ref{eq:dvhub}, the projector derivative term is efficiently computed with an analytical formula~\cite{florisHubbardcorrected2020a} while the occupation-matrix derivative $d n_{m_1 m_2}^{I\sigma}$ is computed with DFPT %, with contributions from the bare derivative of the occupation matrix and the 
and includes contributions from the response of the wave functions to the atomic displacements~\cite{florisHubbardcorrected2020a}:
%\begin{equation} \label{eq:dvdns}
%\sum_{I m_1 m_2}U^I  \hat{P}_{m_1 m_2}^I
%\end{equation}
\begin{equation}\label{eq:dns}
\begin{split}
d n_{m_1 m_2}^{I\sigma}&= \sum_{n\kk}\bra{\psi_{n\kk\sigma}}\partial \hat{P}_{m_2 m_1}^I\ket{\psi_{n\kk\sigma}}  \\
&\!\!\!\!\!\!\!\!\!\!\!\!+\sum_{n\kk} \left[ \bra{  d \psi_{n\kk\sigma}} \hat{P}_{m_2 m_1}^I  \ket{ \psi_{n\kk\sigma} }+\bra{ \psi_{n\kk\sigma}}  \hat{P}_{m_2 m_1}^I \ket{ d \psi_{n\kk\sigma} }   \right]\!\!.
\end{split}
\end{equation}
\indent
% part 4: technical/computational details
We apply this framework to investigate the $e$-ph interactions and electron spectral functions in CoO, focusing on the effects of the Hubbard~\textit{U} correction.  
The ground state electronic structure of CoO is obtained with collinear spin-polarized DFT+\textit{U} calculations in a plane-wave basis using the {\sc Quantum ESPRESSO} code~\cite{giannozziQUANTUM2009}.  
%AFII acronym not defined as implemented in
We use the PBEsol exchange-correlation functional~\cite{perdewRestoring2008} and ultrasoft pseudopotentials~\cite{vanderbiltSoft1990} from the GBRV library~\cite{garrityPseudopotentials2014}.
We employ a four-atoms rhombohedral unit cell~\cite{florisHubbardcorrected2020a} with relaxed lattice constants ($a=5.206$~\r{A}, $b=3.019$~\r{A}, $c=3.009$~\r{A} and angle $\beta=125.05^{\circ}$) and kinetic energy cutoffs of 60~Ry for the wave functions and 720~Ry for the charge density.
Leveraging a recent implementation of DFPT+\textit{U}~\cite{Giannozzi_2017,florisHubbardcorrected2020a}, we compute the lattice dynamics and $e$-ph perturbation potentials %phonon dispersions and eigenvectors
%along with the derivative occupation matrices, $d n_{m_1 m_2}^{I\sigma}$ in Eq.~(\ref{eq:dns}), 
on coarse irreducible $\qq$-point grids. 
We then rotate the KS and Hubbard perturbation potentials with the {\sc Perturbo} code to obtain the $e$-ph matrix elements in the full Brillouin zone (BZ), using coarse grids with $8\times8\times8$ $\kk$ and $\qq$ points. The Wannier functions are obtained with the {\sc Wannier90} code~\cite{pizziWannier902020} and used in {\sc Perturbo}~\cite{zhouPerturbo2020} to interpolate the $e$-ph matrix elements to finer grids. We use atomic orbitals as the basis for the Hubbard manifold. Our method is free of adjustable parameters, including the Hubbard~\textit{U} value, $U=4.55$ eV for Co 3$d$ states, which is determined \textit{ab initio} with a linear response approach~\cite{florisHubbardcorrected2020a, timrovHubbard2018, timrov2021self}. 
%entirely the value of the Hubbard~\textit{U}
\\
\indent
% Self Energy
Using these quantities, we compute the %spin-dependent 
lowest-order $e$-ph self-energy, $\Sigma_{n\kk\sigma}(\omega,T)$, at temperature $T$ and electron energy $\omega$, as implemented in {\sc Perturbo}~\cite{zhouPerturbo2020,zhouPredicting2019}; 
the imaginary part is computed off-shell on a fine energy grid while the real part is evaluated on-shell at the band energy $\varepsilon_{n\kk\sigma}$. 
%\begin{equation}\label{eq:SelfEnergy}
%\begin{split}
%\Sigma_{n\kk\sigma}&(\omega,T)=\frac{2\pi}{\hbar}\sum_{m \nu \qq}\abs{g_{mn\nu}^\sigma(\kk,\qq)}^2\\
%&\times[ \frac{N_{\nu \qq}+f_{m\kk+\qq\sigma}}{\omega-\varepsilon_{m\vec\kk+\qq\sigma}+\omega_{\nu\qq}+i\eta}+\frac{N_{\nu \qq }+1-f_{m\kk+\qq\sigma}}{\omega-\varepsilon_{m\kk+\qq\sigma}-\hbar\omega_{\nu\qq}+i\eta}  ],
%\end{split}
%\end{equation}
%Here, $\varepsilon_{n\kk\sigma}$ and $\omega_{\nu \qq}$ are the electron and phonon energies, respectively, $f_{n\kk\sigma}$ and $N_{\nu \qq}$ the corresponding temperature-dependent occupations, and $\eta$ is a small boradening. 
%
%We obtain the the imaginary part of the self-energy, $\text{Im}\Sigma_{n\kk\sigma}(\omega)$, off-shell on a fine energy $\omega$ grid, and the real part of the self-energy, $\text{Re}\Sigma_{n\kk\sigma}(\omega)$, on-shell at the electron energy $\varepsilon_{n\kk\sigma}$.\\
%\indent
To capture strong $e$-ph interactions beyond the lowest-order, we use the finite-temperature cumulant approach described in Ref.~\cite{zhouPredicting2019}. The latter allows us to obtain the temperature-dependent retarded Green's function $G_{n\kk\sigma}^R(\omega)$ and the resulting electron spectral function, $A_{n\kk\sigma}(\omega)\!=\!-\text{Im}G_{n\kk\sigma}^R(\omega)/\pi$, which includes polaron effects such as band renormalization and satellite peaks~\cite{zhouPredicting2019}. Our framework therefore captures two key aspects of the physics of correlated TMOs, the effects of the localized Coulomb repulsion through DFT$+U$ and the strong $e$-ph coupling and its temperature dependence with the finite-temperature cumulant approach~\cite{zhouPredicting2019}.
\\
\indent
%
%
%\section{Results and discussion}
%
% -------------  FIGURE 1:  E-PH COUPLING  -----------------
%
\begin{figure}[!t]
\includegraphics[width=1.0\columnwidth]{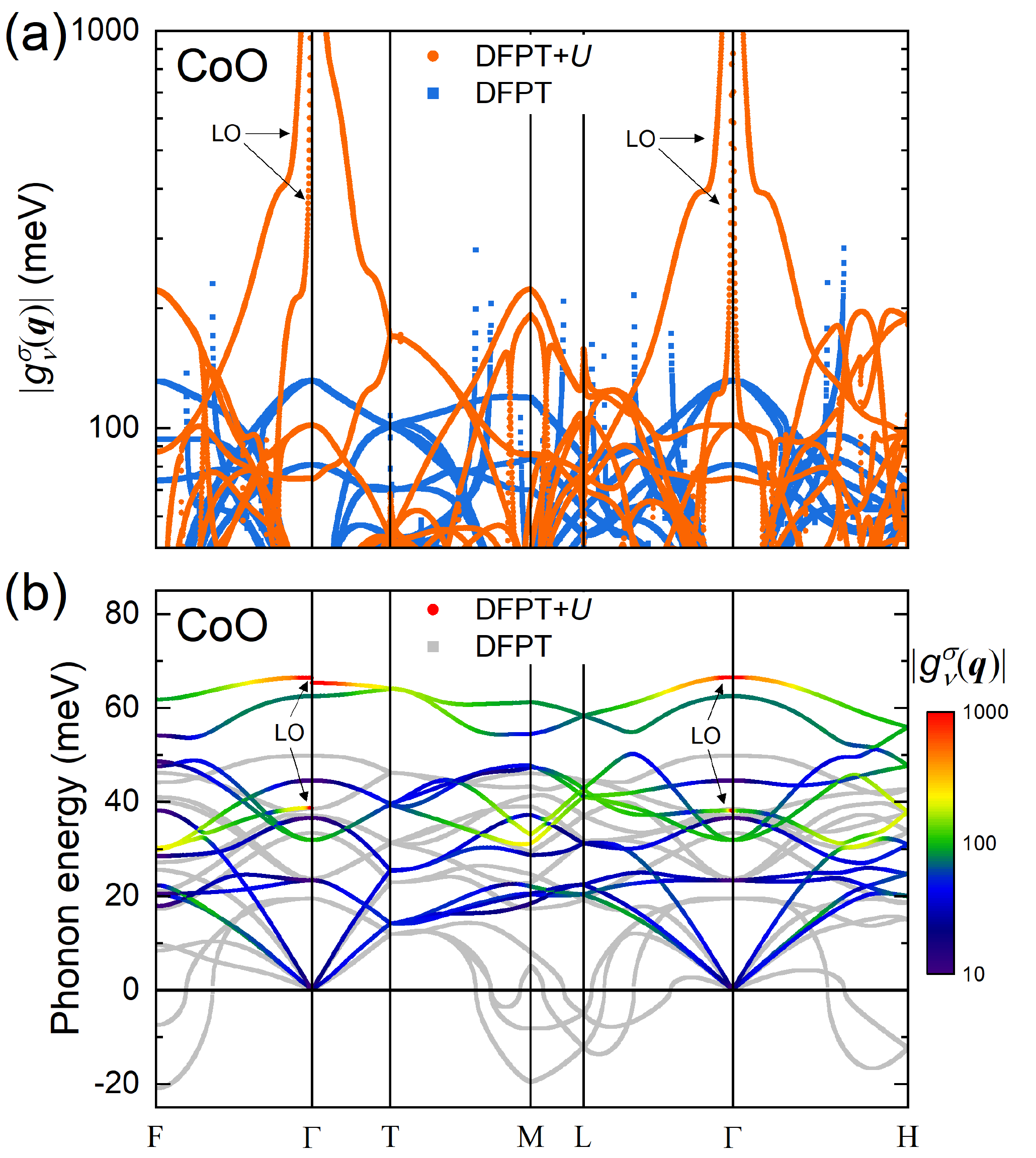}
\caption{(a) Comparison between the $e$-ph matrix elements in CoO computed with DFPT+\textit{U} (orange) and standard DFPT (blue). Shown is $\abs{g_\nu^\sigma(\qq)} \equiv (\sum_{mn} \abs{g_{mn\nu}^\sigma(\kk=0,\qq)}^2/N_b)^{1/2}$ for all phonon modes $\nu$, where the phonon wave-vector $\qq$ is varied along high-symmetry BZ lines and the summation runs over the $N_b=10$ spin-up Co 3\textit{d} bands~\cite{fnote1}. 
%initial electron momentum $\kk$ fixed at the $\Gamma$ point and the
(b) CoO phonon dispersions overlaid with a log-scale color map of $\abs{g_\nu^\sigma(\qq)}$ computed with DFPT+\textit{U} (colored line). The CoO phonon dispersions from standard DFPT (gray line) are given for comparison, with imaginary frequencies shown as negative values.
}\label{fig:Ueph}
\end{figure}
%Here we present the impact of the Hubbard \textit{U} term on the $e$-ph matrix elements, using the spin-up subspace as a demonstration.
The $e$-ph matrix elements from DFPT+\textit{U}, which include effects from both the KS and Hubbard perturbations, are computed for the $d$ bands of CoO in Fig.~\ref{fig:Ueph}(a) and compared with results from standard DFPT~\cite{fnote1}.
The Hubbard \textit{U} correction has a dramatic effect $-$ the two sets of $d$ band $e$-ph matrix elements differ widely, 
for all phonon modes and everywhere in the BZ. 
The largest difference occurs near the zone center, where the DFPT+\textit{U} results show the presence of the Fr\"ohlich interaction~\cite{frohlichElectrons1954}, whereby the $e$-ph matrix elements diverge as $q \!\rightarrow \!0$ for the longitudinal optical (LO) modes~\cite{zhouInitio2016}, whereas in plain DFPT they approach a finite value. 
% with a $\sim$100 meV value
The reason for this difference is subtle $-$ although CoO is a semiconductor with a 2.5 eV band gap~\cite{vanelpElectronic1991}, DFT fails to properly describe its \textit{d} electrons due to self-interaction errors and incorrectly predicts CoO to be a metal, so the Born effective charges and the Fr\"ohlich interaction vanish in DFPT. 
When the Hubbard \textit{U} correction is included, the self-interaction errors are mitigated and CoO is correctly predicted to be a polar semiconductor with divergent $e$-ph coupling for LO phonons near the zone center. This hallmark of the Fr\"ohlich interaction is of critical importance for studies of transport and carrier dynamics in polar materials~\cite{zhouInitio2016,zhouElectronPhonon2018,zhouPredicting2019}. %materials with $d$ electrons. 
\\
\indent
Figure~\ref{fig:Ueph}(b) highlights the dramatic differences in the phonon dispersions computed with DFPT+\textit{U} and plain DFPT~\cite{florisHubbardcorrected2020a}. 
%The two phonon dispersions show major differences in their magnitude and sign. of the system predicted to be 
In the latter, the ground state is dynamically unstable and the phonon dispersions exhibit soft phonon modes with imaginary frequencies.
These errors are propagated to the $e$-ph interactions, resulting in $e$-ph matrix elements with unphysical divergences $-$ near the $F$, $M$, $L$, and $H$ points of the BZ in Fig.~\ref{fig:Ueph}(a) $-$ corresponding to zero-frequency phonon modes [see Eq.~(\ref{eq:me})]~\cite{zhouElectronPhonon2018}. 
%because the $e$-ph matrix elements are inversely proportional to the square root of the phonon frequency. The locations of these divergences in the BZ coincide with 
%semiconducting 
In DFPT+\textit{U}, the ground state is stabilized to the correct antiferromagnetic phase, and the phonon dispersions are significantly improved~\cite{florisHubbardcorrected2020a}; the soft phonon modes are removed entirely and the $e$-ph matrix elements are well behaved throughout the BZ, without spurious divergences.
These results underscore the importance of the Hubbard \textit{U} correction for describing the electronic ground state and the resulting $e$-ph interactions in correlated TMOs.
\\
\indent
%
% -------------  FIGURE 2:  KS vs TOTAL  -----------------
%
\begin{figure}[!t] % strength 
\centering
\includegraphics[width=0.92\columnwidth]{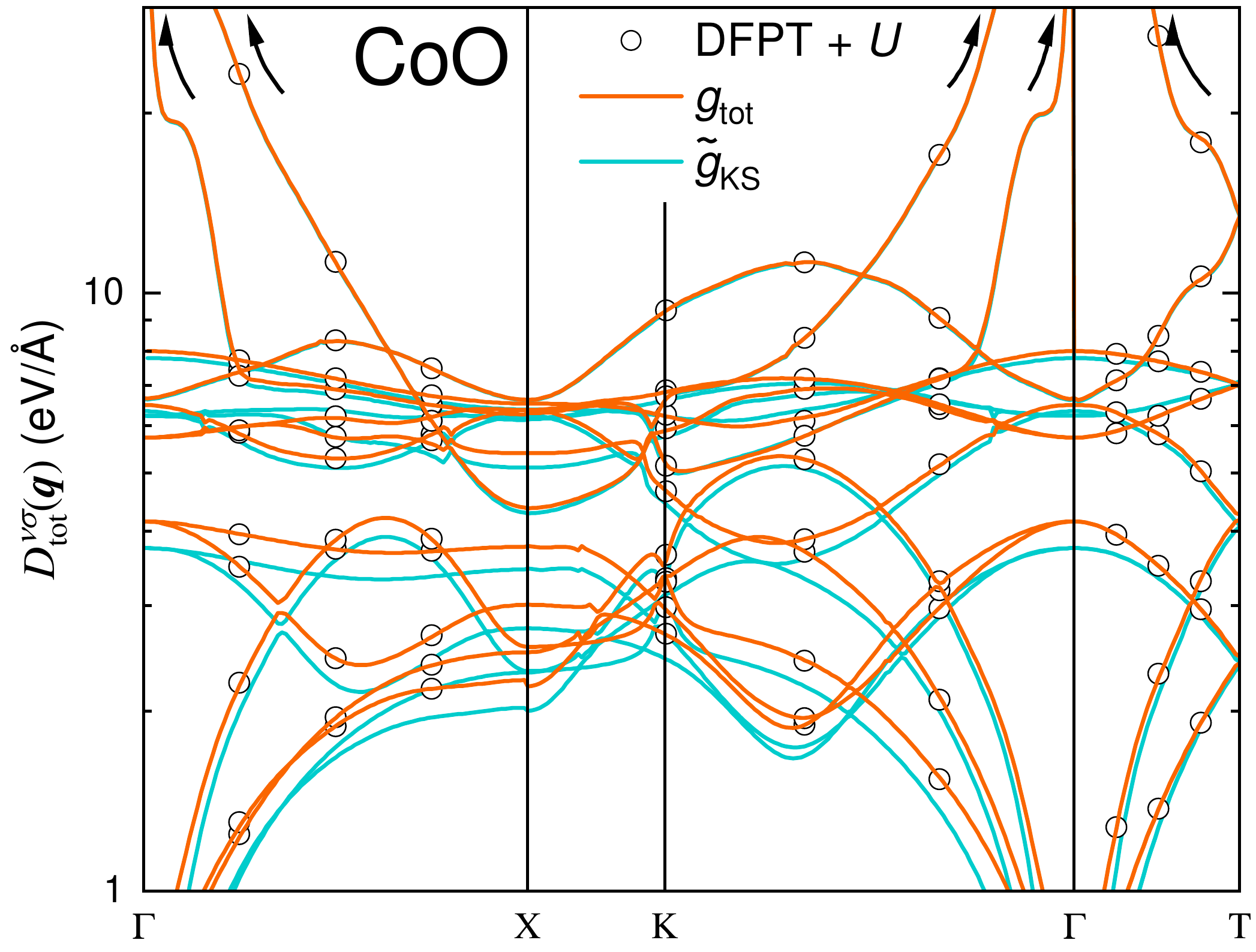}
\caption{
%Gauge-invariant $e$-ph coupling $D_\text{tot}^{\nu\sigma} (\qq)$ in CoO, computed using the spin-up valence bands. %The choice of electron and phonon momenta is the same as in Fig.~\ref{fig:Ueph}. % 
Comparison between the $e$-ph coupling from the KS potential contribution alone (cyan line) and the total result including the Hubbard correction (orange line). 
In each case, we show the gauge-invariant $e$-ph coupling strength~\cite{zhouPerturbo2020}, 
$D_\text{tot}^{\nu\sigma} (\qq) \!=\! (2\omega_{\nu\qq}M_{\rm uc} \sum_{nm} \abs{g_{nm\nu}^\sigma(\kk=0,\qq)}^2/N_b \,)^{1/2}$, 
computed respectively with $e$-ph matrix elements $\tilde{g}_{\rm KS}$ and $g_{\rm tot}$, summing over all $N_b\!=\!13$ spin-up valence bands.   
The BZ labeling refers to an equivalent (distorted) rocksalt structure~\cite{florisHubbardcorrected2020a}. 
%but combines contributions from multiple electronic bands.
%
Direct DFPT$+\,U$ calculations (circles), shown as a benchmark, 
validate the Wannier interpolation. The arrows indicate the divergence due to the Fr\"ohlich interaction.
%DFPT$+\,U$ calculations plus Wannier interpolation (orange line). 
}\label{fig:DFPTU} % The electron momentum $\kk$ is fixed at the $\Gamma$ point and the phonon wave-vector $\qq$ is varied along high-symmetry lines in the Brillouin zone
\end{figure}
In CoO, correcting the wave functions and charge density with DFT+\textit{U} provides the main improvement to the $e$-ph coupling. To illustrate this point, Fig.~\ref{fig:DFPTU} shows that the $e$-ph matrix elements computed with the KS perturbation alone but with DFT+\textit{U} wave functions, $\tilde{g}_{\text {KS}} \propto \bra{\psi_{\rm Hub}} d \hat{V}_{\text{KS}} \ket{\psi_{\rm Hub}}$, can capture both the Fr\"ohlich interaction and the main trends in the $e$-ph coupling. %throughout the BZ. 
Yet, the Hubbard perturbation potential $d{\hat{V}}_{\text{Hub}}$, which describes the effect of the lattice dynamics on the Hubbard \textit{U} correction, also gives an important contribution. %sizable
%\\ 
%\indent
Figure~\ref{fig:DFPTU} compares $\tilde{g}_{\rm KS}$ with the total DFPT+\textit{U} $e$-ph matrix elements, $g_{\rm tot} \propto \bra{\psi_{\rm Hub}}  d \hat{V}_{\text{KS}} + d \hat{V}_{\text{Hub}} \ket{\psi_{\rm Hub}}$, showing that both the KS and Hubbard terms are needed for quantitative accuracy. 
% (still computed with the DFT+\textit{U} wave functions)
%The contribution of the Hubbard~\textit{U} $e$-ph perturbation is  even more apparent if one uses 
Direct DFPT+\textit{U} calculations, shown in Fig.~\ref{fig:DFPTU} as a benchmark, confirm this point and also validate our interpolation procedure. %
%
%The comparison in Fig.~\ref{fig:DFPTU} employs the gauge-invariant $e$-ph coupling strength~\cite{zhouPerturbo2020}, $D_\text{tot}^{\nu\sigma} (\qq) \propto (\, \sum_{nm} \abs{g_{nm\nu}^\sigma(\kk,\qq)}^2 \,)^{1/2}$, which is proportional to the $e$-ph matrix elements but combines contributions from multiple electronic bands. 
% 
\\
\indent
%There are two sets of 3\textit{d} bands in CoO, each generated by a different Co atom in the unit cell; one set of 3\textit{d} bands is fully occupied and the other partially occupied.
Further analysis reveals that the contribution of the Hubbard $e$-ph perturbation  %$\partial{\hat{V}}_{\text{Hub}}$ in Eq.~(\ref{eq:dvhub}), 
is strongly band dependent and acts primarily on the partially filled 3$d$ states of each spin channel~\cite{fnote2}.
%~\footnote{Note that there are two Co atoms in the unit cell of CoO, each generating a set of 3\textit{d} bands in the solid; for each spin channel, due to the antiferromagnetic ground state of CoO, one 3\textit{d} band is completely filled and the other is partially filled.}. 
%
% -------------  FIGURE 3:  DECAY, ROLE OF d BANDS  -----------------
%
\begin{figure}[t!]
\includegraphics[width=1.0\columnwidth]{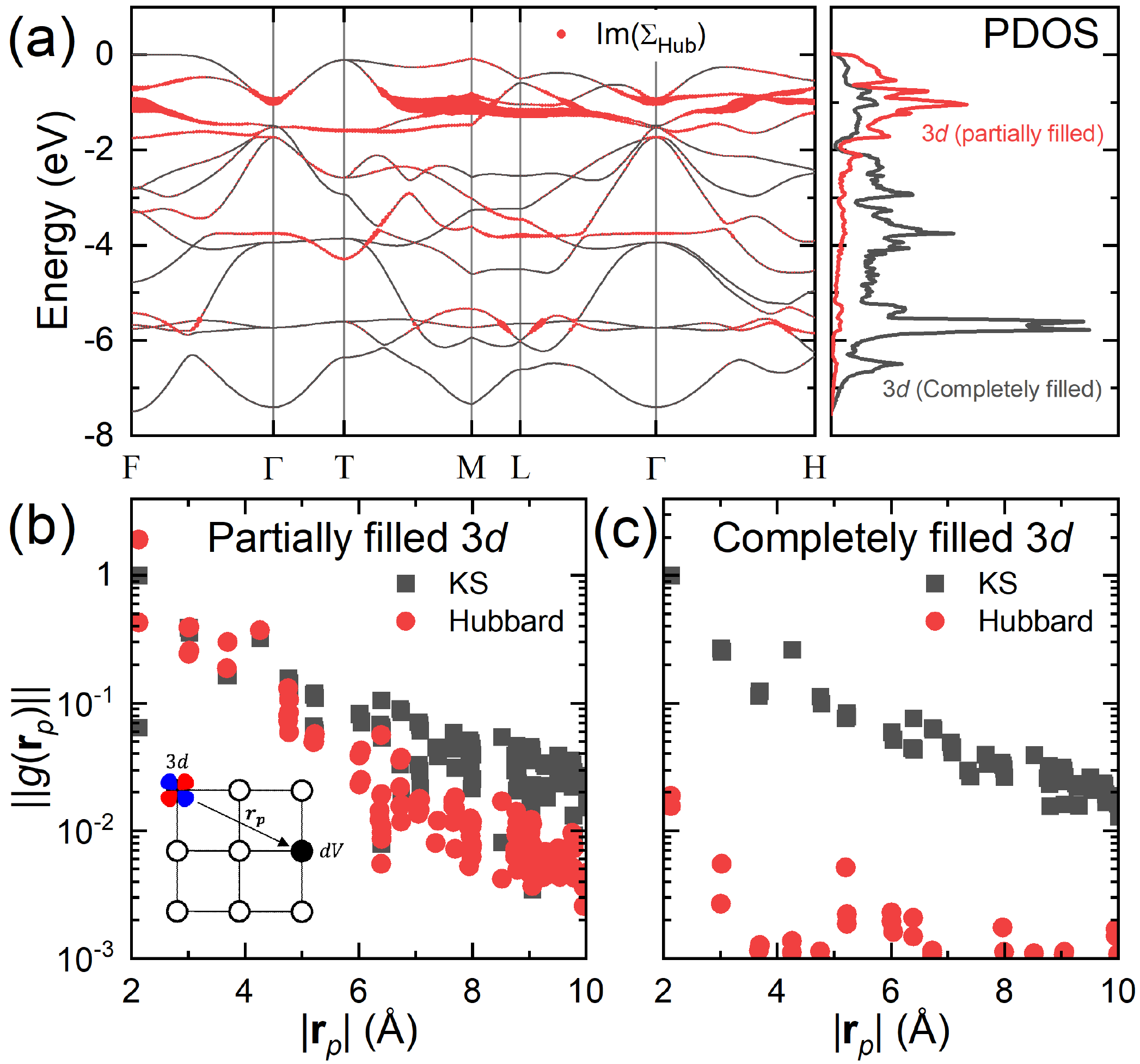}
\caption{(a) Band structure of CoO overlaid with the Hubbard contribution to the imaginary part of the $e$-ph self-energy, $\text{Im}(\Sigma_\text{Hub})$, for the representative case of the spin-up bands. The right panel shows the projected density of states (PDOS) of the partially and completely filled 3$d$ orbitals.
(b),(c) Comparison between the spatial decay of the real-space $e$-ph matrix elements for (b) the partially filled and (c) the completely filled 3$d$ orbitals. 
Shown are the contributions from the KS (black squares) and Hubbard (red circles) $e$-ph perturbations [see Eq.~(\ref{eq:dv})] to the maximum value of the Wannier basis matrix elements~\cite{agapitoInitio2018}, $\abs{\abs{ g(\rr_{p}) }} = \max_{ij} \abs{ g_{ij}(\rr_{p}) }$,  
normalized using the KS contribution. 
The inset in (b) is a schematic of the $e$-ph matrix elements in the Wannier basis, showing the atomic displacement perturbation at distance $|\rr_{p}|$. 
}\label{fig:spatial_decay}
\end{figure}
%
%\indent
%There are two Co atoms in the unit cell of CoO, each generating a set of 3\textit{d} bands in the solid; for each spin channel, due to the antiferromagnetic ground state of CoO one 3\textit{d} band is fully occupied and the other partially occupied. 
%and their 3\textit{d} orbitals (for a fixed spin) are fully occupied for one atom and partially occupied for the other. 
%We find that only electrons in the partially occupied 3\textit{d} bands feel the effect of the Hubbard $e$-ph perturbation.
To demonstrate this result, we compute the imaginary part of the $e$-ph self-energy~\cite{zhouPerturbo2020} with contribution from only the Hubbard $e$-ph perturbation, and map it on the electronic spin-up band structure in Fig.~\ref{fig:spatial_decay}(a). 
The plot shows the selective contribution of the Hubbard perturbation to  $e$-ph processes in the partially filled 3\textit{d} bands and the nearly negligible contribution in the completely filled 3\textit{d} bands. The situation is analogous for the spin-down bands.
\\
\indent
% WF basis e-ph
This trend is confirmed by studying the $e$-ph matrix elements in the Wannier basis, 
$g_{ij}(\rr_{p}) \!\propto\! \bra{\phi_i(0)} d \hat{V}(\rr_{p}) \ket{\phi_j(0)}$~\cite{zhouPerturbo2020}, computed using Co 3\textit{d} Wannier functions $\phi_i$ and $\phi_j$ located on the same Co atom. 
These $e$-ph matrix elements decay exponentially with perturbation distance $|\rr_{p}|$ due to the localized nature of the 3\textit{d} Wannier functions. 
For these local $e$-ph interactions, we find that the KS and Hubbard contributions are nearly identical for the Co atom with partially filled spin-up 3\textit{d} orbitals [Fig.~\ref{fig:spatial_decay}(b)], whereas for the Co atom with completely filled spin-up 3\textit{d} orbitals the Hubbard contribution is orders of magnitude smaller than the KS contribution~[Fig.~\ref{fig:spatial_decay}(c)]~\cite{fnote2}.  
\\
%
% FIGURE 4 -- POLARON SPECTRAL FUNCTION
%
\begin{figure}[t!]
\includegraphics[width=1.0\columnwidth]{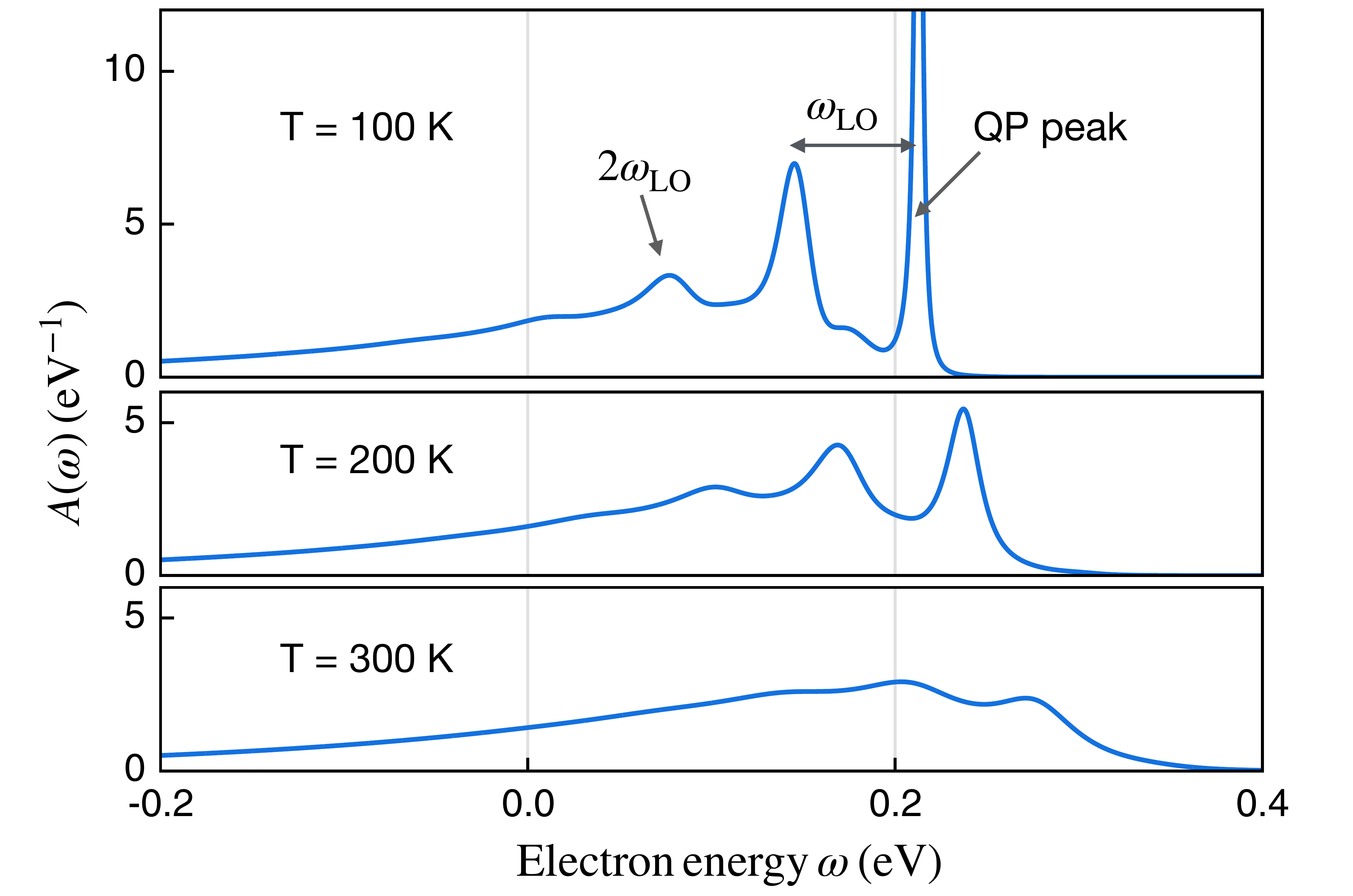}
\caption{Electron spectral function in CoO, computed at three temperatures from 100~K to 300~K, for the highest valence band at crystal momentum $\kk = \mathbf{F}$. In each panel, the zero of the electron energy $\omega$ is set to the DFT+\textit{U} band energy. }
\label{fig:specfunc}
\end{figure}
%
%
%\indent common
%A distinctive feature of $e$-ph interactions in TMOs is that they are often strong enough to form large polaron states due to the polar character of the bonds and the resulting dominant Fr\"ohlich $e$-ph interaction from LO phonons. 
% partially ionic (polar)
\indent
In TMOs, due to the polar bonds, electrons typically couple strongly with LO phonons via the Fr\"ohlich interaction. In this common scenario, the $e$-ph interactions are strong enough to form large polarons, which can dominate transport and electron dynamical processes. 
% but cannot be described with lowest-order perturbation theory~\cite{zhouPredicting2019}. 
%but need to be described accurately electrons 
%these polaron effects is that they are not captured by lowest-order perturbation theory, but they are also not strong enough to 
%
The dominant coupling of electrons with LO phonons is clearly seen in Fig.~\ref{fig:Ueph}(b), and thus we expect significant polaron effects in CoO.  
To investigate them, we compute the electron spectral function with our recently developed finite-temperature cumulant approach, using the DFPT+\textit{U} $e$-ph matrix elements as input~\cite{zhouPredicting2019}. 
\\
\indent
Figure~\ref{fig:specfunc} shows the computed electron spectral functions at three temperatures between 100 $-$ 300~K, for an electronic state near the top of the valence band.  The spectral function at 100~K shows a sharp quasiparticle (QP) peak and two prominent sideband peaks, respectively at energies $\omega_{\rm LO}$ and $2\omega_{\rm LO}$ below the QP peak, where $\omega_{\rm LO} \approx 65$~meV is the energy of the zone center LO phonon with strongest $e$-ph coupling [see Fig.~\ref{fig:Ueph}(b)]. These phonon sidebands are a hallmark of strong $e$-ph coupling and polaron effects~\cite{zhouPredicting2019}. 
% accompanying the QP peak 
Note that our calculations are performed with the Fermi energy lying above the valence band edge $-$ a situation corresponding to lightly $p$-doped CoO $-$ so the QP peak corresponds to a holelike QP excitation. Accordingly, the phonon sidebands appear at energy lower than the QP peak~\cite{Damascelli} and are associated with the simultaneous excitation of a holelike QP plus one or two LO phonons, respectively.
\\
\indent
Due to a well-known sum rule, the spectral function integrates to one over energy, and thus the phonon sidebands transfer spectral weight from the QP peak. %it is noteworthy that 
In CoO, the QP spectral weight is strongly renormalized to a value of 0.2 at 100~K, with significant weight transfer to the phonon sidebands due to the strong $e$-ph interactions.  As the temperature increases from 100 to 200~K, the QP peak becomes broader and overlaps with the phonon sidebands. At 300~K and higher temperatures, the peaks merge into a continuous background and the QP peak representing the original electronic state melts entirely into a polaron excitation. 
%These results demonstrate the strong $e$-ph coupling and polaron effects in CoO. 
As the Fr\"ohlich interaction making up the large polaron is entirely missing in DFT, our study of polaron effects in TMOs is enabled by the correct account of $e$-ph interactions in the DFT+\textit{U} framework developed in this work.
\\
\indent
%\section{Conclusion}
%summary
In summary, we introduced an \textit{ab initio} approach enabling quantitative calculations of $e$-ph interactions and polarons in correlated systems. Our method can be applied broadly to various families of strongly correlated materials with localized \textit{d} or \textit{f} electrons, leveraging the framework of parameter-free DFT+\textit{U}. As shown in this work, our formalism can capture the strong coupling of electron, spin and lattice degrees of freedom in CES and their combined effect on the $e$-ph interactions, paving the way for quantitative studies of the rich physics of various families of strongly correlated materials.
% Our combined treatment of strong $e$-ph coupling and Coulomb interactions at localized $d$ orbitals paves the way for 
%We demonstrate the importance of the Hubbard \textit{U} correction for the $e$-ph coupling of partially filled $d$ bands and for investigating polaron effects in TMOs. 
%Our results indicate e-ph coupling plays an important role on the electronic properties of TMO, and a combined treatment of e-ph coupling and e-e correlation is crucial for TMOs. 
%
%unraveling the combined effect of strong $e$-ph and electron Coulomb interactions in TMOs or strongly correlated materials.
%DFT+\textit{U} as a starting point for e-e + e-ph combined calculations.
\\
\indent

\noindent
Work at Caltech was supported by the National Science Foundation under Grant No. DMR-1750613. J.-J.Z. was supported by the Joint Center for Artificial Photosynthesis, a DOE Energy Innovation Hub, supported through the Office of Science of the U.S. Department of Energy under Award No. DESC0004993. J.P. acknowledges support by the Korea Foundation for Advanced Studies. M.B. was partially supported by the Air Force Office of Scientific Research through the Young Investigator Program Grant No. FA955018-1-0280. 
M.C, I.T. and N.M. acknowledge support from the EU-H2020 NFFA (Grant Agreement No. 654360). I.T. and N.M. also acknowledge support by the Swiss National Science Foundation (SNSF), through Grant No. 200021-179138, and its National Centre of Competence in Research (NCCR) MARVEL. A.F. thanks the UK's HEC Materials Chemistry Consortium, funded by EPSRC (EP/L000202, EP/R029431).
This research used resources of the National Energy Research Scientific Computing Center, a DOE Office of Science User Facility supported by the Office of Science of the U.S. Department of Energy under Contract No. DE-AC02-05CH11231.
\vspace{-5pt}
\bibliographystyle{apsrev4-2}

\begin{thebibliography}{50}%
\makeatletter
\providecommand \@ifxundefined [1]{%
 \@ifx{#1\undefined}
}%
\providecommand \@ifnum [1]{%
 \ifnum #1\expandafter \@firstoftwo
 \else \expandafter \@secondoftwo
 \fi
}%
\providecommand \@ifx [1]{%
 \ifx #1\expandafter \@firstoftwo
 \else \expandafter \@secondoftwo
 \fi
}%
\providecommand \natexlab [1]{#1}%
\providecommand \enquote  [1]{``#1''}%
\providecommand \bibnamefont  [1]{#1}%
\providecommand \bibfnamefont [1]{#1}%
\providecommand \citenamefont [1]{#1}%
\providecommand \href@noop [0]{\@secondoftwo}%
\providecommand \href [0]{\begingroup \@sanitize@url \@href}%
\providecommand \@href[1]{\@@startlink{#1}\@@href}%
\providecommand \@@href[1]{\endgroup#1\@@endlink}%
\providecommand \@sanitize@url [0]{\catcode `\\12\catcode `\$12\catcode
  `\&12\catcode `\#12\catcode `\^12\catcode `\_12\catcode `\%12\relax}%
\providecommand \@@startlink[1]{}%
\providecommand \@@endlink[0]{}%
\providecommand \url  [0]{\begingroup\@sanitize@url \@url }%
\providecommand \@url [1]{\endgroup\@href {#1}{\urlprefix }}%
\providecommand \urlprefix  [0]{URL }%
\providecommand \Eprint [0]{\href }%
\providecommand \doibase [0]{https://doi.org/}%
\providecommand \selectlanguage [0]{\@gobble}%
\providecommand \bibinfo  [0]{\@secondoftwo}%
\providecommand \bibfield  [0]{\@secondoftwo}%
\providecommand \translation [1]{[#1]}%
\providecommand \BibitemOpen [0]{}%
\providecommand \bibitemStop [0]{}%
\providecommand \bibitemNoStop [0]{.\EOS\space}%
\providecommand \EOS [0]{\spacefactor3000\relax}%
\providecommand \BibitemShut  [1]{\csname bibitem#1\endcsname}%
\let\auto@bib@innerbib\@empty
%</preamble>
\bibitem [{\citenamefont {Pickett}(1989)}]{pickettElectronic1989}%
  \BibitemOpen
  \bibfield  {author} {\bibinfo {author} {\bibfnamefont {W.~E.}\ \bibnamefont
  {Pickett}},\ }\href {https://doi.org/10.1103/RevModPhys.61.433} {\bibfield
  {journal} {\bibinfo  {journal} {Rev. Mod. Phys.}\ }\textbf {\bibinfo {volume}
  {61}},\ \bibinfo {pages} {433} (\bibinfo {year} {1989})}\BibitemShut
  {NoStop}%
\bibitem [{\citenamefont {Mott}(1949)}]{mottBasis1949}%
  \BibitemOpen
  \bibfield  {author} {\bibinfo {author} {\bibfnamefont {N.~F.}\ \bibnamefont
  {Mott}},\ }\href {https://doi.org/10.1088/0370-1298/62/7/303} {\bibfield
  {journal} {\bibinfo  {journal} {Proc. Phys. Soc. A}\ }\textbf {\bibinfo
  {volume} {62}},\ \bibinfo {pages} {416} (\bibinfo {year} {1949})}\BibitemShut
  {NoStop}%
\bibitem [{\citenamefont {Mott}(1968)}]{mottMetalInsulator1968}%
  \BibitemOpen
  \bibfield  {author} {\bibinfo {author} {\bibfnamefont {N.~F.}\ \bibnamefont
  {Mott}},\ }\href {https://doi.org/10.1103/RevModPhys.40.677} {\bibfield
  {journal} {\bibinfo  {journal} {Rev. Mod. Phys.}\ }\textbf {\bibinfo {volume}
  {40}},\ \bibinfo {pages} {677} (\bibinfo {year} {1968})}\BibitemShut
  {NoStop}%
\bibitem [{\citenamefont {Ramirez}(1997)}]{ramirezColossal1997}%
  \BibitemOpen
  \bibfield  {author} {\bibinfo {author} {\bibfnamefont {A.~P.}\ \bibnamefont
  {Ramirez}},\ }\href {https://doi.org/10.1088/0953-8984/9/39/005} {\bibfield
  {journal} {\bibinfo  {journal} {J. Phys.: Condens. Matter}\ }\textbf
  {\bibinfo {volume} {9}},\ \bibinfo {pages} {8171} (\bibinfo {year}
  {1997})}\BibitemShut {NoStop}%
\bibitem [{\citenamefont {Schmid}(1994)}]{schmidMultiferroic1994}%
  \BibitemOpen
  \bibfield  {author} {\bibinfo {author} {\bibfnamefont {H.}~\bibnamefont
  {Schmid}},\ }\href {https://doi.org/10.1080/00150199408245120} {\bibfield
  {journal} {\bibinfo  {journal} {Ferroelectrics}\ }\textbf {\bibinfo {volume}
  {162}},\ \bibinfo {pages} {317} (\bibinfo {year} {1994})}\BibitemShut
  {NoStop}%
\bibitem [{\citenamefont {Baroni}\ \emph {et~al.}(2001)\citenamefont {Baroni},
  \citenamefont {{de Gironcoli}}, \citenamefont {Dal~Corso},\ and\
  \citenamefont {Giannozzi}}]{baroniPhonons2001}%
  \BibitemOpen
  \bibfield  {author} {\bibinfo {author} {\bibfnamefont {S.}~\bibnamefont
  {Baroni}}, \bibinfo {author} {\bibfnamefont {S.}~\bibnamefont {{de
  Gironcoli}}}, \bibinfo {author} {\bibfnamefont {A.}~\bibnamefont
  {Dal~Corso}},\ and\ \bibinfo {author} {\bibfnamefont {P.}~\bibnamefont
  {Giannozzi}},\ }\href {https://doi.org/10.1103/RevModPhys.73.515} {\bibfield
  {journal} {\bibinfo  {journal} {Rev. Mod. Phys.}\ }\textbf {\bibinfo {volume}
  {73}},\ \bibinfo {pages} {515} (\bibinfo {year} {2001})}\BibitemShut
  {NoStop}%
\bibitem [{\citenamefont {Bernardi}(2016)}]{bernardiFirstprinciples2016}%
  \BibitemOpen
  \bibfield  {author} {\bibinfo {author} {\bibfnamefont {M.}~\bibnamefont
  {Bernardi}},\ }\href {https://doi.org/10.1140/epjb/e2016-70399-4} {\bibfield
  {journal} {\bibinfo  {journal} {Eur. Phys. J. B}\ }\textbf {\bibinfo {volume}
  {89}},\ \bibinfo {pages} {239} (\bibinfo {year} {2016})}\BibitemShut
  {NoStop}%
\bibitem [{\citenamefont {Agapito}\ and\ \citenamefont
  {Bernardi}(2018)}]{agapitoInitio2018}%
  \BibitemOpen
  \bibfield  {author} {\bibinfo {author} {\bibfnamefont {L.~A.}\ \bibnamefont
  {Agapito}}\ and\ \bibinfo {author} {\bibfnamefont {M.}~\bibnamefont
  {Bernardi}},\ }\href {https://doi.org/10.1103/PhysRevB.97.235146} {\bibfield
  {journal} {\bibinfo  {journal} {Phys. Rev. B}\ }\textbf {\bibinfo {volume}
  {97}},\ \bibinfo {pages} {235146} (\bibinfo {year} {2018})}\BibitemShut
  {NoStop}%
\bibitem [{\citenamefont {Bernardi}\ \emph {et~al.}(2014)\citenamefont
  {Bernardi}, \citenamefont {{Vigil-Fowler}}, \citenamefont {Lischner},
  \citenamefont {Neaton},\ and\ \citenamefont {Louie}}]{bernardiInitio2014}%
  \BibitemOpen
  \bibfield  {author} {\bibinfo {author} {\bibfnamefont {M.}~\bibnamefont
  {Bernardi}}, \bibinfo {author} {\bibfnamefont {D.}~\bibnamefont
  {{Vigil-Fowler}}}, \bibinfo {author} {\bibfnamefont {J.}~\bibnamefont
  {Lischner}}, \bibinfo {author} {\bibfnamefont {J.~B.}\ \bibnamefont
  {Neaton}},\ and\ \bibinfo {author} {\bibfnamefont {S.~G.}\ \bibnamefont
  {Louie}},\ }\href {https://doi.org/10.1103/PhysRevLett.112.257402} {\bibfield
   {journal} {\bibinfo  {journal} {Phys. Rev. Lett.}\ }\textbf {\bibinfo
  {volume} {112}},\ \bibinfo {pages} {257402} (\bibinfo {year}
  {2014})}\BibitemShut {NoStop}%
\bibitem [{\citenamefont {Zhou}\ and\ \citenamefont
  {Bernardi}(2016)}]{zhouInitio2016}%
  \BibitemOpen
  \bibfield  {author} {\bibinfo {author} {\bibfnamefont {J.-J.}\ \bibnamefont
  {Zhou}}\ and\ \bibinfo {author} {\bibfnamefont {M.}~\bibnamefont
  {Bernardi}},\ }\href {https://doi.org/10.1103/PhysRevB.94.201201} {\bibfield
  {journal} {\bibinfo  {journal} {Phys. Rev. B}\ }\textbf {\bibinfo {volume}
  {94}},\ \bibinfo {pages} {201201(R)} (\bibinfo {year} {2016})}\BibitemShut
  {NoStop}%
\bibitem [{\citenamefont {Zhou}\ \emph {et~al.}(2018)\citenamefont {Zhou},
  \citenamefont {Hellman},\ and\ \citenamefont
  {Bernardi}}]{zhouElectronPhonon2018}%
  \BibitemOpen
  \bibfield  {author} {\bibinfo {author} {\bibfnamefont {J.-J.}\ \bibnamefont
  {Zhou}}, \bibinfo {author} {\bibfnamefont {O.}~\bibnamefont {Hellman}},\ and\
  \bibinfo {author} {\bibfnamefont {M.}~\bibnamefont {Bernardi}},\ }\href
  {https://doi.org/10.1103/PhysRevLett.121.226603} {\bibfield  {journal}
  {\bibinfo  {journal} {Phys. Rev. Lett.}\ }\textbf {\bibinfo {volume} {121}},\
  \bibinfo {pages} {226603} (\bibinfo {year} {2018})}\BibitemShut {NoStop}%
\bibitem [{\citenamefont {Zhou}\ and\ \citenamefont
  {Bernardi}(2019)}]{zhouPredicting2019}%
  \BibitemOpen
  \bibfield  {author} {\bibinfo {author} {\bibfnamefont {J.-J.}\ \bibnamefont
  {Zhou}}\ and\ \bibinfo {author} {\bibfnamefont {M.}~\bibnamefont
  {Bernardi}},\ }\href {https://doi.org/10.1103/PhysRevResearch.1.033138}
  {\bibfield  {journal} {\bibinfo  {journal} {Phys. Rev. Research}\ }\textbf
  {\bibinfo {volume} {1}},\ \bibinfo {pages} {033138} (\bibinfo {year}
  {2019})}\BibitemShut {NoStop}%
\bibitem [{\citenamefont {Park}\ \emph {et~al.}(2007)\citenamefont {Park},
  \citenamefont {Giustino}, \citenamefont {Cohen},\ and\ \citenamefont
  {Louie}}]{Park-PRL2007}%
  \BibitemOpen
  \bibfield  {author} {\bibinfo {author} {\bibfnamefont {C.-H.}\ \bibnamefont
  {Park}}, \bibinfo {author} {\bibfnamefont {F.}~\bibnamefont {Giustino}},
  \bibinfo {author} {\bibfnamefont {M.~L.}\ \bibnamefont {Cohen}},\ and\
  \bibinfo {author} {\bibfnamefont {S.~G.}\ \bibnamefont {Louie}},\ }\href
  {https://link.aps.org/doi/10.1103/PhysRevLett.99.086804} {\bibfield
  {journal} {\bibinfo  {journal} {Phys. Rev. Lett.}\ }\textbf {\bibinfo
  {volume} {99}},\ \bibinfo {pages} {086804} (\bibinfo {year}
  {2007})}\BibitemShut {NoStop}%
\bibitem [{\citenamefont {Floris}\ \emph {et~al.}(2007)\citenamefont {Floris},
  \citenamefont {Sanna}, \citenamefont {Massidda},\ and\ \citenamefont
  {Gross}}]{Floris2007}%
  \BibitemOpen
  \bibfield  {author} {\bibinfo {author} {\bibfnamefont {A.}~\bibnamefont
  {Floris}}, \bibinfo {author} {\bibfnamefont {A.}~\bibnamefont {Sanna}},
  \bibinfo {author} {\bibfnamefont {S.}~\bibnamefont {Massidda}},\ and\
  \bibinfo {author} {\bibfnamefont {E.}~\bibnamefont {Gross}},\ }\href
  {https://journals.aps.org/prb/abstract/10.1103/PhysRevB.75.054508} {\bibfield
   {journal} {\bibinfo  {journal} {Phys. Rev. B}\ }\textbf {\bibinfo {volume}
  {75}},\ \bibinfo {pages} {054508} (\bibinfo {year} {2007})}\BibitemShut
  {NoStop}%
\bibitem [{\citenamefont {Sjakste}\ \emph {et~al.}(2014)\citenamefont
  {Sjakste}, \citenamefont {Timrov}, \citenamefont {Gava}, \citenamefont
  {Mingo},\ and\ \citenamefont {Vast}}]{Sjakste}%
  \BibitemOpen
  \bibfield  {author} {\bibinfo {author} {\bibfnamefont {J.}~\bibnamefont
  {Sjakste}}, \bibinfo {author} {\bibfnamefont {I.}~\bibnamefont {Timrov}},
  \bibinfo {author} {\bibfnamefont {P.}~\bibnamefont {Gava}}, \bibinfo {author}
  {\bibfnamefont {N.}~\bibnamefont {Mingo}},\ and\ \bibinfo {author}
  {\bibfnamefont {N.}~\bibnamefont {Vast}},\ }\href
  {http://dl.begellhouse.com/ru/references/5756967540dd1b03,7deb9f2f1087a9e3,7c2384aa250f0b1f.html}
  {\bibfield  {journal} {\bibinfo  {journal} {Annual Rev. Heat Transfer}\
  }\textbf {\bibinfo {volume} {17}} (\bibinfo {year} {2014})}\BibitemShut
  {NoStop}%
\bibitem [{\citenamefont {Li}(2015)}]{liElectrical2015}%
  \BibitemOpen
  \bibfield  {author} {\bibinfo {author} {\bibfnamefont {W.}~\bibnamefont
  {Li}},\ }\href {https://doi.org/10.1103/PhysRevB.92.075405} {\bibfield
  {journal} {\bibinfo  {journal} {Phys. Rev. B}\ }\textbf {\bibinfo {volume}
  {92}},\ \bibinfo {pages} {075405} (\bibinfo {year} {2015})}\BibitemShut
  {NoStop}%
\bibitem [{\citenamefont {Liu}\ \emph {et~al.}(2017)\citenamefont {Liu},
  \citenamefont {Zhou}, \citenamefont {Liao}, \citenamefont {Singh},\ and\
  \citenamefont {Chen}}]{liuFirstprinciples2017}%
  \BibitemOpen
  \bibfield  {author} {\bibinfo {author} {\bibfnamefont {T.-H.}\ \bibnamefont
  {Liu}}, \bibinfo {author} {\bibfnamefont {J.}~\bibnamefont {Zhou}}, \bibinfo
  {author} {\bibfnamefont {B.}~\bibnamefont {Liao}}, \bibinfo {author}
  {\bibfnamefont {D.~J.}\ \bibnamefont {Singh}},\ and\ \bibinfo {author}
  {\bibfnamefont {G.}~\bibnamefont {Chen}},\ }\href
  {https://doi.org/10.1103/PhysRevB.95.075206} {\bibfield  {journal} {\bibinfo
  {journal} {Phys. Rev. B}\ }\textbf {\bibinfo {volume} {95}},\ \bibinfo
  {pages} {075206} (\bibinfo {year} {2017})}\BibitemShut {NoStop}%
\bibitem [{\citenamefont {Ma}\ \emph {et~al.}(2018)\citenamefont {Ma},
  \citenamefont {Nissimagoudar},\ and\ \citenamefont
  {Li}}]{maFirstprinciples2018}%
  \BibitemOpen
  \bibfield  {author} {\bibinfo {author} {\bibfnamefont {J.}~\bibnamefont
  {Ma}}, \bibinfo {author} {\bibfnamefont {A.~S.}\ \bibnamefont
  {Nissimagoudar}},\ and\ \bibinfo {author} {\bibfnamefont {W.}~\bibnamefont
  {Li}},\ }\href {https://doi.org/10.1103/PhysRevB.97.045201} {\bibfield
  {journal} {\bibinfo  {journal} {Phys. Rev. B}\ }\textbf {\bibinfo {volume}
  {97}},\ \bibinfo {pages} {045201} (\bibinfo {year} {2018})}\BibitemShut
  {NoStop}%
\bibitem [{\citenamefont {Ponc\'e}\ \emph {et~al.}(2018)\citenamefont
  {Ponc\'e}, \citenamefont {Margine},\ and\ \citenamefont
  {Giustino}}]{Ponce-PRB}%
  \BibitemOpen
  \bibfield  {author} {\bibinfo {author} {\bibfnamefont {S.}~\bibnamefont
  {Ponc\'e}}, \bibinfo {author} {\bibfnamefont {E.~R.}\ \bibnamefont
  {Margine}},\ and\ \bibinfo {author} {\bibfnamefont {F.}~\bibnamefont
  {Giustino}},\ }\href {https://link.aps.org/doi/10.1103/PhysRevB.97.121201}
  {\bibfield  {journal} {\bibinfo  {journal} {Phys. Rev. B}\ }\textbf {\bibinfo
  {volume} {97}},\ \bibinfo {pages} {121201} (\bibinfo {year}
  {2018})}\BibitemShut {NoStop}%
\bibitem [{\citenamefont {Park}\ \emph {et~al.}(2020)\citenamefont {Park},
  \citenamefont {Zhou},\ and\ \citenamefont {Bernardi}}]{parkSpinphonon2020}%
  \BibitemOpen
  \bibfield  {author} {\bibinfo {author} {\bibfnamefont {J.}~\bibnamefont
  {Park}}, \bibinfo {author} {\bibfnamefont {J.-J.}\ \bibnamefont {Zhou}},\
  and\ \bibinfo {author} {\bibfnamefont {M.}~\bibnamefont {Bernardi}},\ }\href
  {https://doi.org/10.1103/PhysRevB.101.045202} {\bibfield  {journal} {\bibinfo
   {journal} {Phys. Rev. B}\ }\textbf {\bibinfo {volume} {101}},\ \bibinfo
  {pages} {045202} (\bibinfo {year} {2020})}\BibitemShut {NoStop}%
\bibitem [{\citenamefont {Anisimov}\ \emph {et~al.}(1991)\citenamefont
  {Anisimov}, \citenamefont {Zaanen},\ and\ \citenamefont
  {Andersen}}]{anisimovBand1991}%
  \BibitemOpen
  \bibfield  {author} {\bibinfo {author} {\bibfnamefont {V.~I.}\ \bibnamefont
  {Anisimov}}, \bibinfo {author} {\bibfnamefont {J.}~\bibnamefont {Zaanen}},\
  and\ \bibinfo {author} {\bibfnamefont {O.~K.}\ \bibnamefont {Andersen}},\
  }\href {https://doi.org/10.1103/PhysRevB.44.943} {\bibfield  {journal}
  {\bibinfo  {journal} {Phys. Rev. B}\ }\textbf {\bibinfo {volume} {44}},\
  \bibinfo {pages} {943} (\bibinfo {year} {1991})}\BibitemShut {NoStop}%
\bibitem [{\citenamefont {Anisimov}\ \emph {et~al.}(1996)\citenamefont
  {Anisimov}, \citenamefont {Elfimov}, \citenamefont {Hamada},\ and\
  \citenamefont {Terakura}}]{anisimovChargeordered1996}%
  \BibitemOpen
  \bibfield  {author} {\bibinfo {author} {\bibfnamefont {V.~I.}\ \bibnamefont
  {Anisimov}}, \bibinfo {author} {\bibfnamefont {I.~S.}\ \bibnamefont
  {Elfimov}}, \bibinfo {author} {\bibfnamefont {N.}~\bibnamefont {Hamada}},\
  and\ \bibinfo {author} {\bibfnamefont {K.}~\bibnamefont {Terakura}},\ }\href
  {https://doi.org/10.1103/PhysRevB.54.4387} {\bibfield  {journal} {\bibinfo
  {journal} {Phys. Rev. B}\ }\textbf {\bibinfo {volume} {54}},\ \bibinfo
  {pages} {4387} (\bibinfo {year} {1996})}\BibitemShut {NoStop}%
\bibitem [{\citenamefont {Anisimov}\ \emph {et~al.}(1997)\citenamefont
  {Anisimov}, \citenamefont {Aryasetiawan},\ and\ \citenamefont
  {Lichtenstein}}]{anisimovFirstprinciples1997}%
  \BibitemOpen
  \bibfield  {author} {\bibinfo {author} {\bibfnamefont {V.~I.}\ \bibnamefont
  {Anisimov}}, \bibinfo {author} {\bibfnamefont {F.}~\bibnamefont
  {Aryasetiawan}},\ and\ \bibinfo {author} {\bibfnamefont {A.~I.}\ \bibnamefont
  {Lichtenstein}},\ }\href {https://doi.org/10.1088/0953-8984/9/4/002}
  {\bibfield  {journal} {\bibinfo  {journal} {J. Phys.: Condens. Matter}\
  }\textbf {\bibinfo {volume} {9}},\ \bibinfo {pages} {767} (\bibinfo {year}
  {1997})}\BibitemShut {NoStop}%
\bibitem [{\citenamefont {Dudarev}\ \emph {et~al.}(1998)\citenamefont
  {Dudarev}, \citenamefont {Botton}, \citenamefont {Savrasov}, \citenamefont
  {Humphreys},\ and\ \citenamefont {Sutton}}]{dudarevElectronenergyloss1998}%
  \BibitemOpen
  \bibfield  {author} {\bibinfo {author} {\bibfnamefont {S.~L.}\ \bibnamefont
  {Dudarev}}, \bibinfo {author} {\bibfnamefont {G.~A.}\ \bibnamefont {Botton}},
  \bibinfo {author} {\bibfnamefont {S.~Y.}\ \bibnamefont {Savrasov}}, \bibinfo
  {author} {\bibfnamefont {C.~J.}\ \bibnamefont {Humphreys}},\ and\ \bibinfo
  {author} {\bibfnamefont {A.~P.}\ \bibnamefont {Sutton}},\ }\href
  {https://doi.org/10.1103/PhysRevB.57.1505} {\bibfield  {journal} {\bibinfo
  {journal} {Phys. Rev. B}\ }\textbf {\bibinfo {volume} {57}},\ \bibinfo
  {pages} {1505} (\bibinfo {year} {1998})}\BibitemShut {NoStop}%
\bibitem [{\citenamefont {Heyd}\ \emph {et~al.}(2003)\citenamefont {Heyd},
  \citenamefont {Scuseria},\ and\ \citenamefont {Ernzerhof}}]{heydHybrid2003}%
  \BibitemOpen
  \bibfield  {author} {\bibinfo {author} {\bibfnamefont {J.}~\bibnamefont
  {Heyd}}, \bibinfo {author} {\bibfnamefont {G.~E.}\ \bibnamefont {Scuseria}},\
  and\ \bibinfo {author} {\bibfnamefont {M.}~\bibnamefont {Ernzerhof}},\ }\href
  {https://doi.org/10.1063/1.1564060} {\bibfield  {journal} {\bibinfo
  {journal} {J. Chem. Phys.}\ }\textbf {\bibinfo {volume} {118}},\ \bibinfo
  {pages} {8207} (\bibinfo {year} {2003})}\BibitemShut {NoStop}%
\bibitem [{\citenamefont {Georges}\ \emph {et~al.}(1996)\citenamefont
  {Georges}, \citenamefont {Kotliar}, \citenamefont {Krauth},\ and\
  \citenamefont {Rozenberg}}]{georgesDynamical1996}%
  \BibitemOpen
  \bibfield  {author} {\bibinfo {author} {\bibfnamefont {A.}~\bibnamefont
  {Georges}}, \bibinfo {author} {\bibfnamefont {G.}~\bibnamefont {Kotliar}},
  \bibinfo {author} {\bibfnamefont {W.}~\bibnamefont {Krauth}},\ and\ \bibinfo
  {author} {\bibfnamefont {M.~J.}\ \bibnamefont {Rozenberg}},\ }\href
  {https://doi.org/10.1103/RevModPhys.68.13} {\bibfield  {journal} {\bibinfo
  {journal} {Rev. Mod. Phys.}\ }\textbf {\bibinfo {volume} {68}},\ \bibinfo
  {pages} {13} (\bibinfo {year} {1996})}\BibitemShut {NoStop}%
\bibitem [{\citenamefont {Kotliar}\ \emph {et~al.}(2006)\citenamefont
  {Kotliar}, \citenamefont {Savrasov}, \citenamefont {Haule}, \citenamefont
  {Oudovenko}, \citenamefont {Parcollet},\ and\ \citenamefont
  {Marianetti}}]{kotliarElectronic2006}%
  \BibitemOpen
  \bibfield  {author} {\bibinfo {author} {\bibfnamefont {G.}~\bibnamefont
  {Kotliar}}, \bibinfo {author} {\bibfnamefont {S.~Y.}\ \bibnamefont
  {Savrasov}}, \bibinfo {author} {\bibfnamefont {K.}~\bibnamefont {Haule}},
  \bibinfo {author} {\bibfnamefont {V.~S.}\ \bibnamefont {Oudovenko}}, \bibinfo
  {author} {\bibfnamefont {O.}~\bibnamefont {Parcollet}},\ and\ \bibinfo
  {author} {\bibfnamefont {C.~A.}\ \bibnamefont {Marianetti}},\ }\href
  {https://doi.org/10.1103/RevModPhys.78.865} {\bibfield  {journal} {\bibinfo
  {journal} {Rev. Mod. Phys.}\ }\textbf {\bibinfo {volume} {78}},\ \bibinfo
  {pages} {865} (\bibinfo {year} {2006})}\BibitemShut {NoStop}%
\bibitem [{\citenamefont {Kulik}\ \emph {et~al.}(2006)\citenamefont {Kulik},
  \citenamefont {Cococcioni}, \citenamefont {Scherlis},\ and\ \citenamefont
  {Marzari}}]{Kulik2006}%
  \BibitemOpen
  \bibfield  {author} {\bibinfo {author} {\bibfnamefont {H.~J.}\ \bibnamefont
  {Kulik}}, \bibinfo {author} {\bibfnamefont {M.}~\bibnamefont {Cococcioni}},
  \bibinfo {author} {\bibfnamefont {D.~A.}\ \bibnamefont {Scherlis}},\ and\
  \bibinfo {author} {\bibfnamefont {N.}~\bibnamefont {Marzari}},\ }\href
  {https://journals.aps.org/prl/abstract/10.1103/PhysRevLett.97.103001}
  {\bibfield  {journal} {\bibinfo  {journal} {Phys. Rev. Lett.}\ }\textbf
  {\bibinfo {volume} {97}},\ \bibinfo {pages} {103001} (\bibinfo {year}
  {2006})}\BibitemShut {NoStop}%
\bibitem [{\citenamefont {Kulik}(2015)}]{Kulik2015}%
  \BibitemOpen
  \bibfield  {author} {\bibinfo {author} {\bibfnamefont {H.~J.}\ \bibnamefont
  {Kulik}},\ }\href {https://aip.scitation.org/doi/full/10.1063/1.4922693}
  {\bibfield  {journal} {\bibinfo  {journal} {J. Chem. Phys.}\ }\textbf
  {\bibinfo {volume} {142}},\ \bibinfo {pages} {240901} (\bibinfo {year}
  {2015})}\BibitemShut {NoStop}%
\bibitem [{\citenamefont {Pesant}\ and\ \citenamefont
  {C{\^o}t{\'e}}(2011)}]{pesantDFT2011}%
  \BibitemOpen
  \bibfield  {author} {\bibinfo {author} {\bibfnamefont {S.}~\bibnamefont
  {Pesant}}\ and\ \bibinfo {author} {\bibfnamefont {M.}~\bibnamefont
  {C{\^o}t{\'e}}},\ }\href {https://doi.org/10.1103/PhysRevB.84.085104}
  {\bibfield  {journal} {\bibinfo  {journal} {Phys. Rev. B}\ }\textbf {\bibinfo
  {volume} {84}},\ \bibinfo {pages} {085104} (\bibinfo {year}
  {2011})}\BibitemShut {NoStop}%
\bibitem [{\citenamefont {Baettig}\ \emph {et~al.}(2005)\citenamefont
  {Baettig}, \citenamefont {Ederer},\ and\ \citenamefont
  {Spaldin}}]{baettigFirst2005}%
  \BibitemOpen
  \bibfield  {author} {\bibinfo {author} {\bibfnamefont {P.}~\bibnamefont
  {Baettig}}, \bibinfo {author} {\bibfnamefont {C.}~\bibnamefont {Ederer}},\
  and\ \bibinfo {author} {\bibfnamefont {N.~A.}\ \bibnamefont {Spaldin}},\
  }\href {https://doi.org/10.1103/PhysRevB.72.214105} {\bibfield  {journal}
  {\bibinfo  {journal} {Phys. Rev. B}\ }\textbf {\bibinfo {volume} {72}},\
  \bibinfo {pages} {214105} (\bibinfo {year} {2005})}\BibitemShut {NoStop}%
\bibitem [{\citenamefont {Rondinelli}\ \emph {et~al.}(2009)\citenamefont
  {Rondinelli}, \citenamefont {Eidelson},\ and\ \citenamefont
  {Spaldin}}]{rondinelliNon2009}%
  \BibitemOpen
  \bibfield  {author} {\bibinfo {author} {\bibfnamefont {J.~M.}\ \bibnamefont
  {Rondinelli}}, \bibinfo {author} {\bibfnamefont {A.~S.}\ \bibnamefont
  {Eidelson}},\ and\ \bibinfo {author} {\bibfnamefont {N.~A.}\ \bibnamefont
  {Spaldin}},\ }\href {https://doi.org/10.1103/PhysRevB.79.205119} {\bibfield
  {journal} {\bibinfo  {journal} {Phys. Rev. B}\ }\textbf {\bibinfo {volume}
  {79}},\ \bibinfo {pages} {205119} (\bibinfo {year} {2009})}\BibitemShut
  {NoStop}%
\bibitem [{\citenamefont {Floris}\ \emph {et~al.}(2011)\citenamefont {Floris},
  \citenamefont {{de Gironcoli}}, \citenamefont {Gross},\ and\ \citenamefont
  {Cococcioni}}]{florisVibrational2011}%
  \BibitemOpen
  \bibfield  {author} {\bibinfo {author} {\bibfnamefont {A.}~\bibnamefont
  {Floris}}, \bibinfo {author} {\bibfnamefont {S.}~\bibnamefont {{de
  Gironcoli}}}, \bibinfo {author} {\bibfnamefont {E.~K.~U.}\ \bibnamefont
  {Gross}},\ and\ \bibinfo {author} {\bibfnamefont {M.}~\bibnamefont
  {Cococcioni}},\ }\href {https://doi.org/10.1103/PhysRevB.84.161102}
  {\bibfield  {journal} {\bibinfo  {journal} {Phys. Rev. B}\ }\textbf {\bibinfo
  {volume} {84}},\ \bibinfo {pages} {161102} (\bibinfo {year}
  {2011})}\BibitemShut {NoStop}%
\bibitem [{\citenamefont {Floris}\ \emph {et~al.}(2020)\citenamefont {Floris},
  \citenamefont {Timrov}, \citenamefont {Himmetoglu}, \citenamefont {Marzari},
  \citenamefont {{de Gironcoli}},\ and\ \citenamefont
  {Cococcioni}}]{florisHubbardcorrected2020a}%
  \BibitemOpen
  \bibfield  {author} {\bibinfo {author} {\bibfnamefont {A.}~\bibnamefont
  {Floris}}, \bibinfo {author} {\bibfnamefont {I.}~\bibnamefont {Timrov}},
  \bibinfo {author} {\bibfnamefont {B.}~\bibnamefont {Himmetoglu}}, \bibinfo
  {author} {\bibfnamefont {N.}~\bibnamefont {Marzari}}, \bibinfo {author}
  {\bibfnamefont {S.}~\bibnamefont {{de Gironcoli}}},\ and\ \bibinfo {author}
  {\bibfnamefont {M.}~\bibnamefont {Cococcioni}},\ }\href
  {https://doi.org/10.1103/PhysRevB.101.064305} {\bibfield  {journal} {\bibinfo
   {journal} {Phys. Rev. B}\ }\textbf {\bibinfo {volume} {101}},\ \bibinfo
  {pages} {064305} (\bibinfo {year} {2020})}\BibitemShut {NoStop}%
\bibitem [{\citenamefont {Blanchard}\ \emph {et~al.}(2014)\citenamefont
  {Blanchard}, \citenamefont {Balan}, \citenamefont {Giura}, \citenamefont
  {B{\'e}neut}, \citenamefont {Yi}, \citenamefont {Morin}, \citenamefont
  {Pinilla}, \citenamefont {Lazzeri},\ and\ \citenamefont
  {Floris}}]{Blanchard2014}%
  \BibitemOpen
  \bibfield  {author} {\bibinfo {author} {\bibfnamefont {M.}~\bibnamefont
  {Blanchard}}, \bibinfo {author} {\bibfnamefont {E.}~\bibnamefont {Balan}},
  \bibinfo {author} {\bibfnamefont {P.}~\bibnamefont {Giura}}, \bibinfo
  {author} {\bibfnamefont {K.}~\bibnamefont {B{\'e}neut}}, \bibinfo {author}
  {\bibfnamefont {H.}~\bibnamefont {Yi}}, \bibinfo {author} {\bibfnamefont
  {G.}~\bibnamefont {Morin}}, \bibinfo {author} {\bibfnamefont
  {C.}~\bibnamefont {Pinilla}}, \bibinfo {author} {\bibfnamefont
  {M.}~\bibnamefont {Lazzeri}},\ and\ \bibinfo {author} {\bibfnamefont
  {A.}~\bibnamefont {Floris}},\ }\href
  {https://link.springer.com/article/10.1007/s00269-013-0648-7} {\bibfield
  {journal} {\bibinfo  {journal} {Phys. Chem. Miner.}\ }\textbf {\bibinfo
  {volume} {41}},\ \bibinfo {pages} {289} (\bibinfo {year} {2014})}\BibitemShut
  {NoStop}%
\bibitem [{\citenamefont {Miwa}(2018)}]{miwaPrediction2018}%
  \BibitemOpen
  \bibfield  {author} {\bibinfo {author} {\bibfnamefont {K.}~\bibnamefont
  {Miwa}},\ }\href {https://doi.org/10.1103/PhysRevB.97.075143} {\bibfield
  {journal} {\bibinfo  {journal} {Phys. Rev. B}\ }\textbf {\bibinfo {volume}
  {97}},\ \bibinfo {pages} {075143} (\bibinfo {year} {2018})}\BibitemShut
  {NoStop}%
\bibitem [{\citenamefont {Timrov}\ \emph {et~al.}(2018)\citenamefont {Timrov},
  \citenamefont {Marzari},\ and\ \citenamefont
  {Cococcioni}}]{timrovHubbard2018}%
  \BibitemOpen
  \bibfield  {author} {\bibinfo {author} {\bibfnamefont {I.}~\bibnamefont
  {Timrov}}, \bibinfo {author} {\bibfnamefont {N.}~\bibnamefont {Marzari}},\
  and\ \bibinfo {author} {\bibfnamefont {M.}~\bibnamefont {Cococcioni}},\
  }\href {https://doi.org/10.1103/PhysRevB.98.085127} {\bibfield  {journal}
  {\bibinfo  {journal} {Phys. Rev. B}\ }\textbf {\bibinfo {volume} {98}},\
  \bibinfo {pages} {085127} (\bibinfo {year} {2018})}\BibitemShut {NoStop}%
\bibitem [{\citenamefont {Zhou}\ \emph {et~al.}()\citenamefont {Zhou},
  \citenamefont {Park}, \citenamefont {Lu}, \citenamefont {Maliyov},
  \citenamefont {Tong},\ and\ \citenamefont {Bernardi}}]{zhouPerturbo2020}%
  \BibitemOpen
  \bibfield  {author} {\bibinfo {author} {\bibfnamefont {J.-J.}\ \bibnamefont
  {Zhou}}, \bibinfo {author} {\bibfnamefont {J.}~\bibnamefont {Park}}, \bibinfo
  {author} {\bibfnamefont {I.-T.}\ \bibnamefont {Lu}}, \bibinfo {author}
  {\bibfnamefont {I.}~\bibnamefont {Maliyov}}, \bibinfo {author} {\bibfnamefont
  {X.}~\bibnamefont {Tong}},\ and\ \bibinfo {author} {\bibfnamefont
  {M.}~\bibnamefont {Bernardi}},\ }\href {https://arxiv.org/abs/2002.02045}
  {\bibinfo  {journal} {arXiv 2002.02045}\ }\BibitemShut {NoStop}%
\bibitem [{\citenamefont {Giannozzi}\ \emph {et~al.}(2017)\citenamefont
  {Giannozzi}, \citenamefont {Andreussi}, \citenamefont {Brumme}, \citenamefont
  {Bunau}, \citenamefont {Nardelli}, \citenamefont {Calandra}, \citenamefont
  {Car}, \citenamefont {Cavazzoni}, \citenamefont {Ceresoli}, \citenamefont
  {Cococcioni}, \citenamefont {Colonna}, \citenamefont {Carnimeo},
  \citenamefont {Corso}, \citenamefont {de~Gironcoli}, \citenamefont {Delugas},
  \citenamefont {DiStasio}, \citenamefont {Ferretti}, \citenamefont {Floris},
  \citenamefont {Fratesi}, \citenamefont {Fugallo}, \citenamefont {Gebauer},
  \citenamefont {Gerstmann}, \citenamefont {Giustino}, \citenamefont {Gorni},
  \citenamefont {Jia}, \citenamefont {Kawamura}, \citenamefont {Ko},
  \citenamefont {Kokalj}, \citenamefont {Kü{\c{c}}ükbenli}, \citenamefont
  {Lazzeri}, \citenamefont {Marsili}, \citenamefont {Marzari}, \citenamefont
  {Mauri}, \citenamefont {Nguyen}, \citenamefont {Nguyen}, \citenamefont {de-la
  Roza}, \citenamefont {Paulatto}, \citenamefont {Ponc{\'{e}}}, \citenamefont
  {Rocca}, \citenamefont {Sabatini}, \citenamefont {Santra}, \citenamefont
  {Schlipf}, \citenamefont {Seitsonen}, \citenamefont {Smogunov}, \citenamefont
  {Timrov}, \citenamefont {Thonhauser}, \citenamefont {Umari}, \citenamefont
  {Vast}, \citenamefont {Wu},\ and\ \citenamefont {Baroni}}]{Giannozzi_2017}%
  \BibitemOpen
\bibfield  {journal} {  }\bibfield  {author} {\bibinfo {author} {\bibfnamefont
  {P.}~\bibnamefont {Giannozzi}}, \bibinfo {author} {\bibfnamefont
  {O.}~\bibnamefont {Andreussi}}, \bibinfo {author} {\bibfnamefont
  {T.}~\bibnamefont {Brumme}}, \bibinfo {author} {\bibfnamefont
  {O.}~\bibnamefont {Bunau}}, \bibinfo {author} {\bibfnamefont {M.~B.}\
  \bibnamefont {Nardelli}}, \bibinfo {author} {\bibfnamefont {M.}~\bibnamefont
  {Calandra}}, \bibinfo {author} {\bibfnamefont {R.}~\bibnamefont {Car}},
  \bibinfo {author} {\bibfnamefont {C.}~\bibnamefont {Cavazzoni}}, \bibinfo
  {author} {\bibfnamefont {D.}~\bibnamefont {Ceresoli}}, \bibinfo {author}
  {\bibfnamefont {M.}~\bibnamefont {Cococcioni}}, \bibinfo {author}
  {\bibfnamefont {N.}~\bibnamefont {Colonna}}, \bibinfo {author} {\bibfnamefont
  {I.}~\bibnamefont {Carnimeo}}, \bibinfo {author} {\bibfnamefont {A.~D.}\
  \bibnamefont {Corso}}, \bibinfo {author} {\bibfnamefont {S.}~\bibnamefont
  {de~Gironcoli}}, \bibinfo {author} {\bibfnamefont {P.}~\bibnamefont
  {Delugas}}, \bibinfo {author} {\bibfnamefont {R.~A.}\ \bibnamefont
  {DiStasio}}, \bibinfo {author} {\bibfnamefont {A.}~\bibnamefont {Ferretti}},
  \bibinfo {author} {\bibfnamefont {A.}~\bibnamefont {Floris}}, \bibinfo
  {author} {\bibfnamefont {G.}~\bibnamefont {Fratesi}}, \bibinfo {author}
  {\bibfnamefont {G.}~\bibnamefont {Fugallo}}, \bibinfo {author} {\bibfnamefont
  {R.}~\bibnamefont {Gebauer}}, \bibinfo {author} {\bibfnamefont
  {U.}~\bibnamefont {Gerstmann}}, \bibinfo {author} {\bibfnamefont
  {F.}~\bibnamefont {Giustino}}, \bibinfo {author} {\bibfnamefont
  {T.}~\bibnamefont {Gorni}}, \bibinfo {author} {\bibfnamefont
  {J.}~\bibnamefont {Jia}}, \bibinfo {author} {\bibfnamefont {M.}~\bibnamefont
  {Kawamura}}, \bibinfo {author} {\bibfnamefont {H.-Y.}\ \bibnamefont {Ko}},
  \bibinfo {author} {\bibfnamefont {A.}~\bibnamefont {Kokalj}}, \bibinfo
  {author} {\bibfnamefont {E.}~\bibnamefont {Kü{\c{c}}ükbenli}}, \bibinfo
  {author} {\bibfnamefont {M.}~\bibnamefont {Lazzeri}}, \bibinfo {author}
  {\bibfnamefont {M.}~\bibnamefont {Marsili}}, \bibinfo {author} {\bibfnamefont
  {N.}~\bibnamefont {Marzari}}, \bibinfo {author} {\bibfnamefont
  {F.}~\bibnamefont {Mauri}}, \bibinfo {author} {\bibfnamefont {N.~L.}\
  \bibnamefont {Nguyen}}, \bibinfo {author} {\bibfnamefont {H.-V.}\
  \bibnamefont {Nguyen}}, \bibinfo {author} {\bibfnamefont {A.~O.}\
  \bibnamefont {de-la Roza}}, \bibinfo {author} {\bibfnamefont
  {L.}~\bibnamefont {Paulatto}}, \bibinfo {author} {\bibfnamefont
  {S.}~\bibnamefont {Ponc{\'{e}}}}, \bibinfo {author} {\bibfnamefont
  {D.}~\bibnamefont {Rocca}}, \bibinfo {author} {\bibfnamefont
  {R.}~\bibnamefont {Sabatini}}, \bibinfo {author} {\bibfnamefont
  {B.}~\bibnamefont {Santra}}, \bibinfo {author} {\bibfnamefont
  {M.}~\bibnamefont {Schlipf}}, \bibinfo {author} {\bibfnamefont {A.~P.}\
  \bibnamefont {Seitsonen}}, \bibinfo {author} {\bibfnamefont {A.}~\bibnamefont
  {Smogunov}}, \bibinfo {author} {\bibfnamefont {I.}~\bibnamefont {Timrov}},
  \bibinfo {author} {\bibfnamefont {T.}~\bibnamefont {Thonhauser}}, \bibinfo
  {author} {\bibfnamefont {P.}~\bibnamefont {Umari}}, \bibinfo {author}
  {\bibfnamefont {N.}~\bibnamefont {Vast}}, \bibinfo {author} {\bibfnamefont
  {X.}~\bibnamefont {Wu}},\ and\ \bibinfo {author} {\bibfnamefont
  {S.}~\bibnamefont {Baroni}},\ }\href
  {https://doi.org/10.1088/1361-648x/aa8f79} {\bibfield  {journal} {\bibinfo
  {journal} {J. Phys.: Condens. Matter}\ }\textbf {\bibinfo {volume} {29}},\
  \bibinfo {pages} {465901} (\bibinfo {year} {2017})}\BibitemShut {NoStop}%
\bibitem [{\citenamefont {Vanderbilt}(1990)}]{vanderbiltSoft1990}%
  \BibitemOpen
  \bibfield  {author} {\bibinfo {author} {\bibfnamefont {D.}~\bibnamefont
  {Vanderbilt}},\ }\href {https://doi.org/10.1103/PhysRevB.41.7892} {\bibfield
  {journal} {\bibinfo  {journal} {Phys. Rev. B}\ }\textbf {\bibinfo {volume}
  {41}},\ \bibinfo {pages} {7892} (\bibinfo {year} {1990})}\BibitemShut
  {NoStop}%
\bibitem [{\citenamefont {Giannozzi}\ \emph {et~al.}(2009)\citenamefont
  {Giannozzi}, \citenamefont {Baroni}, \citenamefont {Bonini}, \citenamefont
  {Calandra}, \citenamefont {Car}, \citenamefont {Cavazzoni}, \citenamefont
  {Ceresoli}, \citenamefont {Chiarotti}, \citenamefont {Cococcioni},
  \citenamefont {Dabo}, \citenamefont {Corso}, \citenamefont {de~Gironcoli},
  \citenamefont {Fabris}, \citenamefont {Fratesi}, \citenamefont {Gebauer},
  \citenamefont {Gerstmann}, \citenamefont {Gougoussis}, \citenamefont
  {Kokalj}, \citenamefont {Lazzeri}, \citenamefont {{Martin-Samos}},
  \citenamefont {Marzari}, \citenamefont {Mauri}, \citenamefont {Mazzarello},
  \citenamefont {Paolini}, \citenamefont {Pasquarello}, \citenamefont
  {Paulatto}, \citenamefont {Sbraccia}, \citenamefont {Scandolo}, \citenamefont
  {Sclauzero}, \citenamefont {Seitsonen}, \citenamefont {Smogunov},
  \citenamefont {Umari},\ and\ \citenamefont
  {Wentzcovitch}}]{giannozziQUANTUM2009}%
  \BibitemOpen
  \bibfield  {author} {\bibinfo {author} {\bibfnamefont {P.}~\bibnamefont
  {Giannozzi}}, \bibinfo {author} {\bibfnamefont {S.}~\bibnamefont {Baroni}},
  \bibinfo {author} {\bibfnamefont {N.}~\bibnamefont {Bonini}}, \bibinfo
  {author} {\bibfnamefont {M.}~\bibnamefont {Calandra}}, \bibinfo {author}
  {\bibfnamefont {R.}~\bibnamefont {Car}}, \bibinfo {author} {\bibfnamefont
  {C.}~\bibnamefont {Cavazzoni}}, \bibinfo {author} {\bibfnamefont
  {D.}~\bibnamefont {Ceresoli}}, \bibinfo {author} {\bibfnamefont {G.~L.}\
  \bibnamefont {Chiarotti}}, \bibinfo {author} {\bibfnamefont {M.}~\bibnamefont
  {Cococcioni}}, \bibinfo {author} {\bibfnamefont {I.}~\bibnamefont {Dabo}},
  \bibinfo {author} {\bibfnamefont {A.~D.}\ \bibnamefont {Corso}}, \bibinfo
  {author} {\bibfnamefont {S.}~\bibnamefont {de~Gironcoli}}, \bibinfo {author}
  {\bibfnamefont {S.}~\bibnamefont {Fabris}}, \bibinfo {author} {\bibfnamefont
  {G.}~\bibnamefont {Fratesi}}, \bibinfo {author} {\bibfnamefont
  {R.}~\bibnamefont {Gebauer}}, \bibinfo {author} {\bibfnamefont
  {U.}~\bibnamefont {Gerstmann}}, \bibinfo {author} {\bibfnamefont
  {C.}~\bibnamefont {Gougoussis}}, \bibinfo {author} {\bibfnamefont
  {A.}~\bibnamefont {Kokalj}}, \bibinfo {author} {\bibfnamefont
  {M.}~\bibnamefont {Lazzeri}}, \bibinfo {author} {\bibfnamefont
  {L.}~\bibnamefont {{Martin-Samos}}}, \bibinfo {author} {\bibfnamefont
  {N.}~\bibnamefont {Marzari}}, \bibinfo {author} {\bibfnamefont
  {F.}~\bibnamefont {Mauri}}, \bibinfo {author} {\bibfnamefont
  {R.}~\bibnamefont {Mazzarello}}, \bibinfo {author} {\bibfnamefont
  {S.}~\bibnamefont {Paolini}}, \bibinfo {author} {\bibfnamefont
  {A.}~\bibnamefont {Pasquarello}}, \bibinfo {author} {\bibfnamefont
  {L.}~\bibnamefont {Paulatto}}, \bibinfo {author} {\bibfnamefont
  {C.}~\bibnamefont {Sbraccia}}, \bibinfo {author} {\bibfnamefont
  {S.}~\bibnamefont {Scandolo}}, \bibinfo {author} {\bibfnamefont
  {G.}~\bibnamefont {Sclauzero}}, \bibinfo {author} {\bibfnamefont {A.~P.}\
  \bibnamefont {Seitsonen}}, \bibinfo {author} {\bibfnamefont {A.}~\bibnamefont
  {Smogunov}}, \bibinfo {author} {\bibfnamefont {P.}~\bibnamefont {Umari}},\
  and\ \bibinfo {author} {\bibfnamefont {R.~M.}\ \bibnamefont {Wentzcovitch}},\
  }\href {https://doi.org/10.1088/0953-8984/21/39/395502} {\bibfield  {journal}
  {\bibinfo  {journal} {J. Phys.: Condens. Matter}\ }\textbf {\bibinfo {volume}
  {21}},\ \bibinfo {pages} {395502} (\bibinfo {year} {2009})}\BibitemShut
  {NoStop}%
\bibitem [{\citenamefont {Perdew}\ \emph {et~al.}(2008)\citenamefont {Perdew},
  \citenamefont {Ruzsinszky}, \citenamefont {Csonka}, \citenamefont {Vydrov},
  \citenamefont {Scuseria}, \citenamefont {Constantin}, \citenamefont {Zhou},\
  and\ \citenamefont {Burke}}]{perdewRestoring2008}%
  \BibitemOpen
  \bibfield  {author} {\bibinfo {author} {\bibfnamefont {J.~P.}\ \bibnamefont
  {Perdew}}, \bibinfo {author} {\bibfnamefont {A.}~\bibnamefont {Ruzsinszky}},
  \bibinfo {author} {\bibfnamefont {G.~I.}\ \bibnamefont {Csonka}}, \bibinfo
  {author} {\bibfnamefont {O.~A.}\ \bibnamefont {Vydrov}}, \bibinfo {author}
  {\bibfnamefont {G.~E.}\ \bibnamefont {Scuseria}}, \bibinfo {author}
  {\bibfnamefont {L.~A.}\ \bibnamefont {Constantin}}, \bibinfo {author}
  {\bibfnamefont {X.}~\bibnamefont {Zhou}},\ and\ \bibinfo {author}
  {\bibfnamefont {K.}~\bibnamefont {Burke}},\ }\href
  {https://doi.org/10.1103/PhysRevLett.100.136406} {\bibfield  {journal}
  {\bibinfo  {journal} {Phys. Rev. Lett.}\ }\textbf {\bibinfo {volume} {100}},\
  \bibinfo {pages} {136406} (\bibinfo {year} {2008})}\BibitemShut {NoStop}%
\bibitem [{\citenamefont {Garrity}\ \emph {et~al.}(2014)\citenamefont
  {Garrity}, \citenamefont {Bennett}, \citenamefont {Rabe},\ and\ \citenamefont
  {Vanderbilt}}]{garrityPseudopotentials2014}%
  \BibitemOpen
  \bibfield  {author} {\bibinfo {author} {\bibfnamefont {K.~F.}\ \bibnamefont
  {Garrity}}, \bibinfo {author} {\bibfnamefont {J.~W.}\ \bibnamefont
  {Bennett}}, \bibinfo {author} {\bibfnamefont {K.~M.}\ \bibnamefont {Rabe}},\
  and\ \bibinfo {author} {\bibfnamefont {D.}~\bibnamefont {Vanderbilt}},\
  }\href
  {https://www.sciencedirect.com/science/article/abs/pii/S0927025613005077}
  {\bibfield  {journal} {\bibinfo  {journal} {Comput. Mater. Sci.}\ }\textbf
  {\bibinfo {volume} {81}},\ \bibinfo {pages} {446} (\bibinfo {year}
  {2014})}\BibitemShut {NoStop}%
\bibitem [{\citenamefont {Pizzi}\ \emph {et~al.}(2020)\citenamefont {Pizzi},
  \citenamefont {Vitale}, \citenamefont {Arita}, \citenamefont {Bl{\"u}gel},
  \citenamefont {Freimuth}, \citenamefont {G{\'e}ranton}, \citenamefont
  {Gibertini}, \citenamefont {Gresch}, \citenamefont {Johnson}, \citenamefont
  {Koretsune}, \citenamefont {{Iba{\~n}ez-Azpiroz}}, \citenamefont {Lee},
  \citenamefont {Lihm}, \citenamefont {Marchand}, \citenamefont {Marrazzo},
  \citenamefont {Mokrousov}, \citenamefont {Mustafa}, \citenamefont {Nohara},
  \citenamefont {Nomura}, \citenamefont {Paulatto}, \citenamefont {Ponc{\'e}},
  \citenamefont {Ponweiser}, \citenamefont {Qiao}, \citenamefont {Th{\"o}le},
  \citenamefont {Tsirkin}, \citenamefont {Wierzbowska}, \citenamefont
  {Marzari}, \citenamefont {Vanderbilt}, \citenamefont {Souza}, \citenamefont
  {Mostofi},\ and\ \citenamefont {Yates}}]{pizziWannier902020}%
  \BibitemOpen
  \bibfield  {author} {\bibinfo {author} {\bibfnamefont {G.}~\bibnamefont
  {Pizzi}}, \bibinfo {author} {\bibfnamefont {V.}~\bibnamefont {Vitale}},
  \bibinfo {author} {\bibfnamefont {R.}~\bibnamefont {Arita}}, \bibinfo
  {author} {\bibfnamefont {S.}~\bibnamefont {Bl{\"u}gel}}, \bibinfo {author}
  {\bibfnamefont {F.}~\bibnamefont {Freimuth}}, \bibinfo {author}
  {\bibfnamefont {G.}~\bibnamefont {G{\'e}ranton}}, \bibinfo {author}
  {\bibfnamefont {M.}~\bibnamefont {Gibertini}}, \bibinfo {author}
  {\bibfnamefont {D.}~\bibnamefont {Gresch}}, \bibinfo {author} {\bibfnamefont
  {C.}~\bibnamefont {Johnson}}, \bibinfo {author} {\bibfnamefont
  {T.}~\bibnamefont {Koretsune}}, \bibinfo {author} {\bibfnamefont
  {J.}~\bibnamefont {{Iba{\~n}ez-Azpiroz}}}, \bibinfo {author} {\bibfnamefont
  {H.}~\bibnamefont {Lee}}, \bibinfo {author} {\bibfnamefont {J.-M.}\
  \bibnamefont {Lihm}}, \bibinfo {author} {\bibfnamefont {D.}~\bibnamefont
  {Marchand}}, \bibinfo {author} {\bibfnamefont {A.}~\bibnamefont {Marrazzo}},
  \bibinfo {author} {\bibfnamefont {Y.}~\bibnamefont {Mokrousov}}, \bibinfo
  {author} {\bibfnamefont {J.~I.}\ \bibnamefont {Mustafa}}, \bibinfo {author}
  {\bibfnamefont {Y.}~\bibnamefont {Nohara}}, \bibinfo {author} {\bibfnamefont
  {Y.}~\bibnamefont {Nomura}}, \bibinfo {author} {\bibfnamefont
  {L.}~\bibnamefont {Paulatto}}, \bibinfo {author} {\bibfnamefont
  {S.}~\bibnamefont {Ponc{\'e}}}, \bibinfo {author} {\bibfnamefont
  {T.}~\bibnamefont {Ponweiser}}, \bibinfo {author} {\bibfnamefont
  {J.}~\bibnamefont {Qiao}}, \bibinfo {author} {\bibfnamefont {F.}~\bibnamefont
  {Th{\"o}le}}, \bibinfo {author} {\bibfnamefont {S.~S.}\ \bibnamefont
  {Tsirkin}}, \bibinfo {author} {\bibfnamefont {M.}~\bibnamefont
  {Wierzbowska}}, \bibinfo {author} {\bibfnamefont {N.}~\bibnamefont
  {Marzari}}, \bibinfo {author} {\bibfnamefont {D.}~\bibnamefont {Vanderbilt}},
  \bibinfo {author} {\bibfnamefont {I.}~\bibnamefont {Souza}}, \bibinfo
  {author} {\bibfnamefont {A.~A.}\ \bibnamefont {Mostofi}},\ and\ \bibinfo
  {author} {\bibfnamefont {J.~R.}\ \bibnamefont {Yates}},\ }\href
  {https://doi.org/10.1088/1361-648X/ab51ff} {\bibfield  {journal} {\bibinfo
  {journal} {J. Phys.: Condens. Matter}\ }\textbf {\bibinfo {volume} {32}},\
  \bibinfo {pages} {165902} (\bibinfo {year} {2020})}\BibitemShut {NoStop}%
\bibitem [{\citenamefont {Timrov}\ \emph {et~al.}(2021)\citenamefont {Timrov},
  \citenamefont {Marzari},\ and\ \citenamefont {Cococcioni}}]{timrov2021self}%
  \BibitemOpen
  \bibfield  {author} {\bibinfo {author} {\bibfnamefont {I.}~\bibnamefont
  {Timrov}}, \bibinfo {author} {\bibfnamefont {N.}~\bibnamefont {Marzari}},\
  and\ \bibinfo {author} {\bibfnamefont {M.}~\bibnamefont {Cococcioni}},\
  }\href {https://journals.aps.org/prb/abstract/10.1103/PhysRevB.103.045141}
  {\bibfield  {journal} {\bibinfo  {journal} {Phys. Rev. B}\ }\textbf {\bibinfo
  {volume} {103}},\ \bibinfo {pages} {045141} (\bibinfo {year}
  {2021})}\BibitemShut {NoStop}%
\bibitem [{fno({\natexlab{a}})}]{fnote1}%
  \BibitemOpen
  \href@noop {} {}\bibinfo {note} {{A}s the Hubbard \textit{U} correction
  reorders the band energies, to make a meaningful comparison in
  Fig.~\ref{fig:Ueph}(a) we average the $e$-ph coupling over the entire set of
  10 $3d$ bands, for both the DFPT and DFPT+\textit{U}
  calculations.}\BibitemShut {Stop}%
\bibitem [{\citenamefont {Fr{\"o}hlich}(1954)}]{frohlichElectrons1954}%
  \BibitemOpen
  \bibfield  {author} {\bibinfo {author} {\bibfnamefont {H.}~\bibnamefont
  {Fr{\"o}hlich}},\ }\href {https://doi.org/10.1080/00018735400101213}
  {\bibfield  {journal} {\bibinfo  {journal} {Adv. Phys.}\ }\textbf {\bibinfo
  {volume} {3}},\ \bibinfo {pages} {325} (\bibinfo {year} {1954})}\BibitemShut
  {NoStop}%
\bibitem [{\citenamefont {{van Elp}}\ \emph {et~al.}(1991)\citenamefont {{van
  Elp}}, \citenamefont {Wieland}, \citenamefont {Eskes}, \citenamefont
  {Kuiper}, \citenamefont {Sawatzky}, \citenamefont {{de Groot}},\ and\
  \citenamefont {Turner}}]{vanelpElectronic1991}%
  \BibitemOpen
  \bibfield  {author} {\bibinfo {author} {\bibfnamefont {J.}~\bibnamefont {{van
  Elp}}}, \bibinfo {author} {\bibfnamefont {J.~L.}\ \bibnamefont {Wieland}},
  \bibinfo {author} {\bibfnamefont {H.}~\bibnamefont {Eskes}}, \bibinfo
  {author} {\bibfnamefont {P.}~\bibnamefont {Kuiper}}, \bibinfo {author}
  {\bibfnamefont {G.~A.}\ \bibnamefont {Sawatzky}}, \bibinfo {author}
  {\bibfnamefont {F.~M.~F.}\ \bibnamefont {{de Groot}}},\ and\ \bibinfo
  {author} {\bibfnamefont {T.~S.}\ \bibnamefont {Turner}},\ }\href
  {https://doi.org/10.1103/PhysRevB.44.6090} {\bibfield  {journal} {\bibinfo
  {journal} {Phys. Rev. B}\ }\textbf {\bibinfo {volume} {44}},\ \bibinfo
  {pages} {6090} (\bibinfo {year} {1991})}\BibitemShut {NoStop}%
\bibitem [{fno({\natexlab{b}})}]{fnote2}%
  \BibitemOpen
  \href@noop {} {}\bibinfo {note} {{N}ote that there are two Co atoms in the
  unit cell of CoO, each generating a set of 3\textit{d} bands in the solid;
  for each spin channel, due to the antiferromagnetic ground state of CoO, one
  set of 3\textit{d} bands is completely filled and the other is partially
  filled.}\BibitemShut {Stop}%
\bibitem [{\citenamefont {Damascelli}\ \emph {et~al.}(2003)\citenamefont
  {Damascelli}, \citenamefont {Hussain},\ and\ \citenamefont
  {Shen}}]{Damascelli}%
  \BibitemOpen
  \bibfield  {author} {\bibinfo {author} {\bibfnamefont {A.}~\bibnamefont
  {Damascelli}}, \bibinfo {author} {\bibfnamefont {Z.}~\bibnamefont
  {Hussain}},\ and\ \bibinfo {author} {\bibfnamefont {Z.-X.}\ \bibnamefont
  {Shen}},\ }\href {https://link.aps.org/doi/10.1103/RevModPhys.75.473}
  {\bibfield  {journal} {\bibinfo  {journal} {Rev. Mod. Phys.}\ }\textbf
  {\bibinfo {volume} {75}},\ \bibinfo {pages} {473} (\bibinfo {year}
  {2003})}\BibitemShut {NoStop}%
\end{thebibliography}

\end{document}